\documentclass[twocolumn, amsmath]{aastex702}
\usepackage{amsmath}

\usepackage{pstricks}

\usepackage{soul}

\begin{document}

\title{A comparison of two- and one-fluid dust solvers for dusty, supersonic turbulence}

\author{Narges Vadood}
\affiliation{Memorial University of Newfoundland, Department of Computer Science, St.\ John’s, NL A1B 3X5, Canada}
\email{nvadood@mun.ca}  

\author[orcid=0000-0002-6238-9096]{Terrence S.\ Tricco} 
\affiliation{Memorial University of Newfoundland, Department of Computer Science, St.\ John’s, NL A1B 3X5, Canada}
\affiliation{Memorial University of Newfoundland, Department of Physics \& Physical Oceanography, St.\ John’s, NL A1B 3X5, Canada}
\email[show]{tstricco@mun.ca}

\begin{abstract}
We model 3 and 10 $\mu$m dust grains in supersonic, turbulent molecular cloud conditions to compare two different numerical methods solving dust coupled to gas through a drag term. One method models dust and gas as two separate species (two fluid) and the other the combination of dust and gas as a single mixture (one fluid) using the diffusion approximation. Simulations are performed in a 3D periodic box using the Phantom code. The gas probability distributions are consistent with a lognormal distribution for both methods and grain sizes. The dust distributions are different in the two methods, showing discrepancies particularly in regions of low density. The most significant difference between the two methods is in the dust-to-gas ratio distributions. Both methods peak at the mean dust-to-gas ratio of 0.01, but the two-fluid method has wider distributions than the one-fluid method suggesting excess dust concentration in dense filaments. Filaments are where the one-fluid method results are most accurate. The narrowing of the one-fluid distribution is also caused by the stopping time limiter used to maintain the terminal velocity approximation. This artificially makes the mixture more coupled in low-density regions. Our overall conclusion is that both methods are viable for the study of dust in molecular clouds, but that the correct method should be chosen based on the numerical Stokes number regime of the calculation to avoid numerical artifacts.
\end{abstract}

%% Keywords should appear after the \end{abstract} command. 
%% The AAS Journals now uses Unified Astronomy Thesaurus concepts:
%% https://astrothesaurus.org
%% You will be asked to selected these concepts during the submission process
%% but this old "keyword" functionality is maintained in case authors want
%% to include these concepts in their preprints.

%% From the front matter, we move on to the body of the paper.
%% Sections are demarcated by \section and \subsection, respectively.
%% Observe the use of the LaTeX \label
%% command after the \subsection to give a symbolic KEY to the
%% subsection for cross-referencing in a \ref command.
%% You can use LaTeX's \ref and \label commands to keep track of
%% cross-references to sections, equations, tables, and figures.
%% That way, if you change the order of any elements, LaTeX will
%% automatically renumber them.
%%
%% We recommend that authors also use the natbib \citep
%% and \citet commands to identify citations.  The citations are
%% tied to the reference list via symbolic KEYs. The KEY corresponds
%% to the KEY in the \bibitem in the reference list below. 

\section{Introduction}

Dust plays a crucial role in the dense interstellar medium (ISM) of our Galaxy, serving as the primary source of opacity and light scattering. Dense molecular clouds are regions in space with relatively high densities of molecular hydrogen compared to other types of ISM, with mean densities of $n_{H_2}\sim$ 100--1000 cm$^{-3}$ \citep{sm06, ballesteros-paredes-etal20, colman-etal24}. These clouds have relatively low temperatures, usually around 10 K, and can span large regions, sometimes hundreds of parsecs across. 

Dense molecular clouds play a significant role in the life cycle of the ISM, acting as a reservoir of molecular gas that can eventually collapse under gravity to form new stars in the denser regions. These clouds are primarily composed of molecular gas (99\%), and most of that is molecular hydrogen, which is difficult to detect directly \citep{habart-etal05}. The second most abundant element is helium, which is a noble gas and invisible. The second most abundant molecule is CO, which has rotational transitions that are easily excited at typical molecular cloud temperatures and densities, making it useful as a tracer of hydrogen in cold molecular clouds to investigate molecular cloud properties \citep{sss79, bwl13}.

These clouds also contain solid dust grains composed of silicates and carbonaceous materials of sizes from a few microns to a few nanometers  \citep{mrn77}, which account for about 1\% of the mass in molecular clouds \citep{bsd78}. For a review on interstellar dust grain properties, see \citet{draine03}. Historically, interstellar dust was first recognized for its ability to obscure starlight. Subsequently, dust was used as a tracer for studying cold molecular clouds \citep{lada-etal94}, providing more detailed and robust measurements of cloud sizes, masses, and internal structures compared to CO observations of the same regions within the Milky Way \citep{gps09}. The advantage of using dust over molecular lines lies in the fact that dust grains do not require a minimum density to be detectable and remain effective at high densities, whereas molecular lines can become problematic.

Molecular clouds exhibit high turbulence \citep{larson81, solomon-etal87, hf12}. Regardless of the underlying mechanisms driving it, turbulence can lead to the accumulation of gas in filaments and cause the formation of complex, filamentary shapes, which resemble the intricate, irregular forms observed in molecular clouds \citep{schneider-etal11}. It is a question on whether turbulence can drive variations in the local dust-to-gas ratio. If so, then this would have implications on derived cloud masses, as the calibration of H$_2$ column density to integrated CO intensity (X factor) is often derived using dust extinction \citep{pcg08, kong-etal15, liseau-etal15, ld20, lld22}.

Several numerical studies have investigated the dynamics of dust grains within cold, turbulent molecular clouds, revealing variations in dust concentration. \citet{hl16} use meshless finite mass \citep[MFM;][]{hopkins15} where two separate fluids of gas and dust are modeled using two types of particles, and they report a variation in dust concentration by more than a factor of 1000 for dust grains larger than 0.01 $\mu$m with their simulation setup. They observed the emergence of dust filaments even in the absence of gas filaments. In \citet{mattsson-etal19}, the gas is considered on a uniform Cartesian grid (Eulerian approach), while the dust fluid is handled using inertial particles (Lagrangian approach). \citet{mattsson-etal19} report large dust-to-gas ratio variations for grain sizes larger than 1 $\mu$m, and \citet{mfv19} show small-scale clustering (increasing the local grain density by at least a factor of a few) for nanometer-sized grains. These findings indicate that dust grains decouple from the gas at these sizes and no longer trace the gas. 

\citet{tpl17} only report significant decoupling for dust grains larger than 10 $\mu$m, with 0.1 $\mu$m dust grains remaining highly coupled to the gas distribution. This is contrast to the results found by \citet{hl16} and \citet{mattsson-etal19}. The key difference between these results is that \citet{tpl17} track the evolution of the barycenter of mass between the gas and the dust using the dust diffusion approximation in smoothed particle hydrodynamics (SPH) simulations. The full one-fluid system of equations governing gas and dust dynamics is mathematically equivalent to the set of equations in the two-fluid formalism, but involves a change of variables that allows tracking the evolution of the mixture's barycenter, relative velocity difference, and dust concentration. This has been further simplified using the diffusion approximation \citep{pl15}, which is grounded in the terminal velocity approximation \citep{yg05, ly17, lp19}. 

While suitable for well-coupled dust grains in molecular clouds (small grains in dense environments), the one-fluid diffusion approximation is inadequate for larger grains as it underestimates the velocity difference between the dust and gas. This means that numerical artifacts may appear for large dust grains or in low-density environments where the dust stopping time exceeds the local dynamical time.  The issue with the two-fluid results, where gas and dust are modeled separately, is that the dust is pressureless. \cite{pf10} demonstrated that pressureless tracer particles in simulations of supersonic turbulence fail to accurately represent dynamics, resulting in numerical artifacts. These artifacts manifest as exaggerated concentration in high-density regions and almost complete absence in underdense regions. Moreover, \cite{lp12a} established that simulations of two-fluid dust and gas mixtures require sufficient gas resolution to adequately resolve the length scale over which dust grains begin comoving with the gas (dust ``stopping length'').

What is needed is a comparative study where the same simulation is performed side by side using the two- and one-fluid dust methods to determine the conditions in which the two methods agree on their solutions for molecular clouds. \citet{commercon-etal23} have recently performed this type of study using the grid-based code Ramses \citep{ramses}. For the two-fluid method, the dust is treated as Lagrangian particles while the gas dynamics are computed on an Eulerian grid. They find similar results as previous studies. For two-fluid simulations, they find artificial dust trapping in high-density regions, with the maximum dust enrichment affected by the gas resolution. For their one-fluid method with dust diffusion approximation simulations, they find that the terminal velocity approximation is well suited for $<10$ $\mu$m dust grains inside dense filaments, but does not accurately capture dust dynamics in low-density regions. They conclude that the numerical properties of these two methods needs further study.

Our goal is to evaluate the effectiveness of the two- and one-fluid in the dust diffusion approximation numerical methods for the evolution of dust and gas mixtures in supersonic, turbulent molecular clouds. Our work is very similar to that of \citet{commercon-etal23}, with a key difference that we include a limiter on the stopping time for the one-fluid method. \citet{ballabio-etal18} recommend this limiter to ensure the equations of motion remain valid within the terminal velocity approximation, particularly in low-density areas where the stopping time might surpass the dynamical time (Courant time). We specifically choose dust grain sizes and molecular cloud conditions to try to meet the numerical requirements of both methods. This intentionally probes the limit of applicability of both methods; that is, the parameter space in which one method starts becoming more effective than the other. Everything is performed in the Phantom SPH code \citep{phantom}, so that any differences found in our results can be attributed entirely to the choice of dust solver.

The paper is organized as follows. The continuum equations described in Section~\ref{sec:continuum}. The one- and two-fluid numerical methods are discussed in Section~\ref{sec:numerics}. The simulation initial conditions are given in Section~\ref{sec:initcond}, and simulation results are presented in Section~\ref{sec:results}. Conclusions are given in Section~\ref{sec:conclusion}.

\section{Continuum equations}
\label{sec:continuum}

Dust and gas are considered as two continuous fluids. Though the dust is comprised of individual grains, and their individual properties are important, the dust is treated as a continuum because the resolution scale of the simulation is far above that of individual grains. Collisions between dust grains are not modeled. The dynamics of the dust are governed by the drag force from the gas, while the dynamics of the gas are influenced by both the drag force and the gas pressure. In the two-fluid set of equations, the dust and gas are treated as distinct continuous fluids. Each fluid is governed by its own set of equations of motion with additional terms representing the drag force that couples the two components. In contrast, the one-fluid description merges the tightly coupled dust and gas into a single, unified fluid. This is mathematically the same, but views the combined mixture of gas and dust from a common point of reference. 

\subsection{The two-fluid method}
The equations describing the evolution of gas and
dust mixtures, where dust is treated as a pressureless, inviscid fluid in the continuum limit, are given by

\begin{equation}
    \frac{{\rm d} \rho_{\rm g}}{{\rm d} t} = -\rho_{\rm g} (\nabla \cdot \mathbf{v}_{\rm g}),
	\label{rho-g}
\end{equation}

\begin{equation}
    \frac{{\rm d} \rho_{\rm d}}{{\rm d} t} = -\rho_{\rm d} (\nabla \cdot \mathbf{v}_{\rm d}),
	\label{rho-d}
\end{equation}

\begin{equation}
     \frac{{\rm d} \mathbf{v}_{\rm g}}{{\rm d} t} = 
    \mathbf{f} + \frac{K}{\rho_{\rm g}}(\mathbf{v}_{\rm d} - \mathbf{v}_{\rm g}) - \frac{\mathbf{\nabla} P_{\rm g}}{\rho_{\rm g}} ,
	\label{mv-g}
\end{equation}

\begin{equation}
 \frac{{\rm d} \mathbf{v}_{\rm d}}{{\rm d}t} =
    \mathbf{f} - \frac{K}{\rho_{\rm d}}(\mathbf{v}_{\rm d} - \mathbf{v}_{\rm g}) ,
	\label{mv-d}
\end{equation}
\begin{equation}
 \frac{{\rm d} u}{{\rm d} t}= - \frac{P_{\rm g}}{\rho_{\rm g}}(\mathbf{\nabla}\cdot\mathbf{v}_{\rm g}) + \frac{K}{\rho_{\rm g}}(\mathbf{v}_{\rm d} - \mathbf{v}_{\rm g})^2,
	\label{int-u}
\end{equation}
where ${\rm d}/{\rm d}t \equiv \partial/\partial t + \mathbf{v} \cdot \nabla$ is the material (Lagrangian) derivative, the subscripts ${\rm g}$ and ${\rm d}$ refer to the gas and dust, respectively, $\rho$ is the density, $\mathbf{v}$ is the velocity, $P_{\rm g}$ is the thermodynamic gas pressure, $u$ is the internal energy, and $\mathbf{f}$ represents acceleration acting on both components, such as gravitational acceleration or turbulent driving. The coefficient $K$
denotes the drag coefficient between the two phases, with dimensions of mass per unit volume per unit time. 
The gas and dust are coupled through the drag terms, which exchange
momentum and energy between the gas and the dust phases.

\subsection{Equation of state}
The equation set is closed by an equation of state defining the gas pressure. 
For an ideal gas, this is given by
 \begin{equation}
    P_{\rm g} = \frac{\rho_{\rm g} k_{\rm B} T}{\mu m_{\rm H}}
	\label{P-T}
\end{equation}
where $k_{\rm B}$ is Boltzmann's constant, $\mu$ is the mean molecular weight and $m_{\rm H}$ is the mass of a hydrogen atom.
The sound speed $c_{\rm s}$ is given by 
\begin{equation}
    c_{\rm s}^2 = \gamma \frac{P_{\rm g}}{\rho_{\rm g}}.
	\label{sound speed}
\end{equation}
where $\gamma$ is the adiabatic index. These equations lead to an equation between sound speed and temperature:
\begin{equation}
    T = {\frac{\mu m_{\rm H}}{k_{\rm B}}} c_{\rm s}^2.
	\label{T-u}
\end{equation}

In this work, we consider cold, dense molecular clouds for which $\gamma=1$ is a reasonable approximation. In this case, the sound speed is constant, as is the temperature of the cloud. Therefore, the gas pressure is calculated according to
\begin{equation}
    P_{\rm g} = c_{\rm s}^2 \rho_{\rm g} ,
\end{equation}
which means the pressure is proportional to the density.

\subsection{Stopping time}
The stopping time is a characteristic timescale which represents the time required for the drag to adjust the velocity of a dust grain to match the gas velocity. It can be calculated according to
 \begin{equation}
    t_{\rm s}= \frac{\rho_{\rm g} \rho_{\rm d}}{K(\rho_{\rm d} + \rho_{\rm g})}.
	\label{stopping-time1}
\end{equation}
In the Epstein regime, the mean free path of the gas is much larger than the size of the dust grains. This condition typically holds in molecular clouds. Under these circumstances, the stopping time for compact, spherical dust grains is
 \begin{equation}
    t_{\rm s} = \frac{\rho_{\rm grain}s_{\rm grain}}{(\rho_{\rm d} + \rho_{\rm g}) c_{\rm s} f}\sqrt{\frac{\pi \gamma}{8}},
	\label{stopping-time2}
\end{equation}
where $\rho_{\rm grain}$ is the intrinsic density of the dust grains, $s_{\rm grain}$ is the dust grain size, and $f$ is a correction for supersonic motion \citep{kwok75}, given by
 \begin{equation}
    f \equiv \sqrt{1 + \frac{9\pi}{128}\frac{\Delta\mathbf{v}^2}{c_{\rm s}^2}}.
	\label{kwok}
\end{equation}
 
Assuming that all grains have the same intrinsic density, the larger the grain, the longer the stopping time, and consequently, the smaller grains are the most coupled to the gas. 

\subsection{Stokes number}
The degree of coupling between the dust and the gas is characterized by
the Stokes number, which is defined as 
 \begin{equation}
    {\rm St} = \frac{t_{\rm s}}{t_{\rm dyn}},
	\label{eq:stokes}
\end{equation}
where $t_{\rm dyn}$ is the typical dynamical time of the system. For ${\rm St} \ll 1$, the gas and dust are dynamically coupled, which implies $\Delta \mathbf{v}\ll c_{\rm s}$ and $f=1$. 

It is important to make the distinction that different Stokes numbers can be defined based on the choice of the dynamical time. In the case of turbulence simulations, two distinct Stokes numbers arise, depending on whether one considers the turbulent crossing time or the local numerical response time of the system to a perturbation (see Section~\ref{sec:stokes-sph})

\subsection{The one-fluid method}
\label{sec:full-1f}

The gas-dust evolution equations given by Equations~\ref{rho-g}--\ref{int-u} can be reformulated to represent a single fluid moving with the barycentric velocity
\begin{equation}
    \mathbf{v} \equiv \frac{\rho_{\rm g}\mathbf{v}_{\rm g} + \rho_{\rm d} \mathbf{v}_{\rm d}}{\rho_{\rm g} + \rho_{\rm d}}. 
	\label{v_c}
\end{equation}

The differential velocity between the two phases, $\Delta \mathbf{v}$, is defined according to
\begin{equation}
    \Delta \mathbf{v} \equiv \mathbf{v}_{\rm d} - \mathbf{v}_{\rm g}.
	\label{delta-v}
\end{equation}

In this approach, the total density, $\rho = \rho_{\rm g} + \rho_{\rm d}$, the barycentric velocity, the differential velocity, and a prescription for specifying the dust fraction, $\epsilon = \rho_{\rm d} /\rho$, can be used to provide a mathematically equivalent formulation of Equations~(\ref{rho-g})--(\ref{int-u}). By using the identities $\mathbf{v}_{\rm g} = \mathbf{v} - (\rho_{\rm d}/\rho)\Delta \mathbf{v}$ and 
$\mathbf{v}_{\rm d} = \mathbf{v} - (\rho_{\rm g}/\rho)\Delta \mathbf{v}$, Equations~(\ref{rho-g})--(\ref{int-u}) become
\begin{equation}
    \frac{{\rm d} \rho}{{\rm d} t} = -\rho (\nabla \cdot \mathbf{v}),
	\label{d-rho}
\end{equation}

 \begin{equation}
    \frac{{\rm d} \epsilon}{{\rm d} t} = - \frac{1}{\rho}\nabla \cdot [\epsilon(1 - \epsilon)\rho \Delta \mathbf{v}],
	\label{epsilon}
\end{equation}

\begin{equation}
    \frac{{\rm d} \mathbf{v}}{{\rm d} t} = \mathbf{f}- \frac{\nabla P_{\rm g}}{\rho} - \frac{1}{\rho}\nabla \cdot
    [\epsilon(1 - \epsilon)\rho \Delta\mathbf{v}\Delta\mathbf{v}],
	\label{dv}
\end{equation}

 \begin{equation}
    \frac{{\rm d}\Delta\mathbf{v}}{{\rm d} t} = - \frac{\Delta\mathbf{v}}{t_{\rm s}} + 
    \frac{\nabla P_{\rm g}}{(1-\epsilon)\rho} - (\Delta \mathbf{v} \cdot \nabla)\mathbf{v} +
    \frac{1}{2}\nabla[(2\epsilon - 1)\Delta\mathbf{v}^2],
	\label{d-delta-v}
\end{equation}

\begin{equation}
    \frac{{\rm d}u}{{\rm d}t} = -\frac{P_{\rm g}}{(1-\epsilon)\rho}\nabla \cdot 
    (\mathbf{v} - \epsilon \Delta\mathbf{v}) + 
    \epsilon (\Delta\mathbf{v} \cdot \nabla) u + \epsilon \frac{\Delta \mathbf{v}^2}{t_{\rm s}} .
	\label{du}
\end{equation}

The dust-to-gas ratio can be calculated according to $\rho_{\rm d} / \rho_{\rm g} = \epsilon / (1 - \epsilon)$.

\subsubsection{Terminal velocity approximation}
\label{sec:TVA}
The fluid equations in the one-fluid formalism can be simplified when the stopping time, $t_{\rm s}$, is small compared to the typical dynamical time-scale, i.e., when ${\rm St} < 1$. In this regime, the relative velocity between the two phases rapidly approaches zero; that is, the dust begins comoving with the gas on a shorter timescale than the dynamical time \citep{yg05}. We can apply this terminal velocity approximation to ignore the time dependence of the differential velocity between the gas and the dust, simplifying Equation~(\ref{d-delta-v}) to
 \begin{equation}
    \Delta\mathbf{v} = \frac{t_{\rm s}}{1 - \epsilon}\frac{\nabla P}{\rho}.
	\label{terminal-v}
\end{equation}

Using this equation for $\Delta \mathbf{v}$, and neglecting terms of second order in $t_{\rm s}$, Equations~(\ref{d-rho})--(\ref{du}) reduce to
\begin{equation}
    \frac{{\rm d} \rho}{{\rm d} t} = -\rho (\nabla \cdot \mathbf{v}),
	\label{d-rho2}
\end{equation}
 \begin{equation}
    \frac{{\rm d} \epsilon}{{\rm d} t} = - \frac{1}{\rho}\nabla \cdot (\epsilon t_{\rm s} \nabla P_{\rm g}),
	\label{d-epsilon}
\end{equation}
\begin{equation}
    \frac{{\rm d}\mathbf{v}}{{\rm d} t} = \mathbf{f}- \frac{\nabla P_{\rm g}}{\rho},
	\label{dv2}
\end{equation}
\begin{equation}
    \frac{{\rm d}u}{{\rm d}t} = -\frac{P_{\rm g}}{(1-\epsilon)\rho}\nabla \cdot \mathbf{v},
	\label{du2}
\end{equation}
These treat the dust evolution as a diffusion process.

The equations in the terminal velocity approximation have a strong similarity to the usual 
hydrodynamic equations, apart from the additional evolution equation for the dust fraction (Equation~\ref{d-epsilon}). From this point onward, we use `one fluid' to refer to simplified model with the terminal velocity approximation, while reserving `full one fluid' for the model without this approximation as described in Section~\ref{sec:full-1f}.

\section{Numerical method}
\label{sec:numerics}

We use the Phantom SPH code \citep{phantom}. Phantom contains numerical methods to solve the gas-dust equations in both the two- and one-fluid (with the terminal velocity approximation) prescriptions. Both methods include the back-reaction of the dust on the gas.

\subsection{Smoothed particle hydrodynamics}

SPH solves the equations of hydrodynamics in Lagrangian form. The fluid is discretized onto a set of particles of fixed mass that are moved with the local fluid velocity. 

\subsubsection{Smoothing kernels}
\label{sec:kernels}

For defining density in SPH, a smoothing kernel is applied to neighboring particles. The density at a given point is computed by summing up the weighted contributions from surrounding particles within a smoothing length, $h_a$, by using a smoothing kernel 
 \begin{equation}
    W_{ab}(h_a) \equiv \frac{\sigma}{h_a^3}f(q_{ab}),
	\label{weight-function}
\end{equation}
where $a$ and $b$ are particle indices, $\sigma$ is a normalization constant, the factor of $h_a^3$ gives the dimensions of inverse volume, and $f(q_{ab})$ is a dimensionless function of $q_{ab} \equiv \vert\mathbf{r}_a - \mathbf{r}_b\vert/h_a$ with $\mathbf{r}$ the position vector.

In this work, we use the $M_4$ cubic spline given by
 \begin{equation}
  f(q) =
    \begin{cases}
      1 - \frac{3}{2}q^2 + \frac{3}{4}q^3, & 0\leq q <1;\\
      \frac{1}{4}(2-q)^3, & 1\leq q <2;\\
      0 & q\geq 2.
    \end{cases}       
	\label{cubic-dernel}
\end{equation}
The cubic spline has a normalization constant $\sigma = 1/\pi$ in 3D. \citet{pf10} have shown that the results of supersonic turbulence are in agreement between grid and SPH codes when the cubic spline is used. Therefore, there is no expectation that using higher-order kernels is necessary for our work. The smoothing length of a particle is given by
\begin{equation}
    h_a = 1.2 \left ( \frac{m_a}{\rho_a}\right )^{1/3} ,
	\label{h}
\end{equation}
where $m$ is the mass of the particle and the factor 1.2 is an appropriate choice for the cubic spline \citep{da12}.

A ``double-hump'' kernel \citep{lp12a} is used to calculate the drag terms in the two-fluid SPH implementation, defining
\begin{equation}
    D_{ab}(h_a) = \frac{\sigma}{h_a^3} q_{ab}^2 f(q_{ab}) .
\end{equation}
The double-hump kernels were found to reduce errors in the calculation of drag forces by a factor of several hundred. This occurs because the kernel is peaked away from the center, which is beneficial as there is, in general, no contributing particle at the center. In other words, there is no self-contribution as the drag is between pairs of different types of particles (note that this would be problematic for calculating density as the self-contribution is most important).

\subsubsection{Time step constraint}
The time step is constrained to be less than the maximum stable time step according to the Courant condition.
For a given particle, $a$, this is given by
\begin{equation}
    \Delta t_{\rm cour}^a  \leq C_{\rm cour}\frac{h_a}{c_{\rm s}}
	\label{cour}
\end{equation}
where $C_{\rm cour} = 0.3$ by default. Individual particle time stepping is used, such that each particle evolves on its own timescale independently. Each particle is assigned a time step which satisfies Equation~(\ref{cour}) and is a factor of 2 decrement from a maximum time step. The time step limiter of \citet{sm09} is used so that neighboring particles are at most within a factor of 2 of one another.

\subsection{The two-fluid method in Phantom}

In the two-fluid method, the mixture is discretized into
two distinct sets of dust and gas particles. Equations (\ref{rho-g})--(\ref{rho-d}) are discretized with a density summation over neighbors of the same type, giving
\begin{equation}
    \rho_a = \sum_{b} m_b W_{ab}(h_a);~~~~~~~~h_a = 1.2 \left ( \frac{m_a}{\rho_a}\right )^{1/3}
	\label{sph-rho-g}
\end{equation}
 for the density of a gas particle, and 

 \begin{equation}
    \rho_i = \sum _{j} m_j W_{ij}(h_i);~~~~~~~~h_i = 1.2 \left ( \frac{m_i}{\rho_i}\right )^{1/3}
	\label{sph-rho-d}
\end{equation}
for the density of a dust particle.
The indices $a$ and $b$ refer to quantities computed on gas particles and $i$ and $j$ refer to quantities computed on dust particles.

The discretized equation of motion for the gas particles can be obtained by writing the Lagrangian of the system consisting of gas and dust in SPH and using the Euler-Lagrange equations. Equation~(\ref{mv-g}) becomes 
\begin{equation}
\begin{aligned}
\frac{{\rm d}\mathbf{v}_a}{{\rm d}t} =
    & -\sum_b m_b \Bigg [ \frac{P_{{\rm g},a}}{\Omega_a \rho_a^2}\nabla_a W_{ab}(h_a)\\ 
    & \hspace{1cm} + \frac{P_{{\rm g},b}}{\Omega_b \rho_b^2}\nabla_a W_{ab}(h_b)\Bigg ] \\
    & -3\sum_j m_j \frac{\mathbf{v}_{aj}\cdot \hat{\mathbf{r}}_{aj}}{(\rho_a + \rho_j) t_{s,aj}} \hat{\mathbf{r}}_{aj} D_{aj}(h_a),
	\label{dva/dt-p} 
\end{aligned}
\end{equation}
where the first summation is the pressure gradient, the second summation is the drag term, $\Omega_a$ is a term related to the gradient of the smoothing length 
\begin{equation}
    \Omega_a \equiv 1- \frac{\partial h_a}{\partial\rho_a}\sum_b m_b \frac{\partial W_{ab}(h_a)}{\partial h_a} ,
\end{equation}
and $D$ is a double-hump kernel (Section~\ref{sec:kernels}).

Equation~(\ref{mv-d}), the equation of motion for dust particles, is discretized according to
\begin{equation}
   \frac{{\rm d}\mathbf{v}_i}{{\rm d}t} = 
      -3\sum_b m_b \frac{\mathbf{v}_{ib}\cdot \hat{\mathbf{r}}_{ib}}{(\rho_i + \rho_b)t_{{\rm s},ib}} \hat{\mathbf{r}}_{ib} D_{ib}(h_b).
	\label{dvi/dt-drag}
\end{equation}

The internal energy equation (Equation~\ref{int-u}) is discretized as
\begin{equation}
\begin{aligned}
    \frac{{\rm d}u_a}{{\rm d}t} = 
      & \frac{P_{{\rm g},a}}{\rho_a^2 \Omega_a}\sum_b m_b \mathbf{v}_{ab}\cdot\nabla_a W_{ab}(h_a) \\
      & + 3\sum_j m_j \frac{(\mathbf{v}_{aj}\cdot \hat{\mathbf{r}}_{aj})^2}{(\rho_a + \rho_j)t_{{\rm s},aj}} \ D_{aj}(h_a),
	\label{dua/dt-p}
\end{aligned}
\end{equation}
where the two terms are from pressure and drag heating, respectively.

 The stopping time, $t_{{\rm s},ai}$, is defined between a pair of gas and dust particles, given by
\begin{equation}
    t_{{\rm s},ai} = \frac{\rho_a \rho_i}{K_{ai}(\rho_a + \rho_i)},
	\label{eq:2f-sph-ts}
\end{equation}
where $a$ refers to a gas particle and $i$ a dust particle.

In the two-fluid method, the time step of a particle is also constrained by the stopping time according to 
\begin{equation}
    \Delta t_{\rm drag}^a = \min_j (t_{{\rm s},aj}).
	\label{deltaTdrag}
\end{equation}

\citet{lp12a} demonstrate the two-fluid method also has a spatial resolution requirement in order to correctly predict the dust dynamics. This is given by
\begin{equation}
    h \lesssim c_{\rm s} t_{\rm s} .
    \label{eq:2f-resolution}
\end{equation}
This is similar to the dimensionless Dahmk\"{o}hler number in mesh-based codes, which measures whether chemical reactions occur within a cell before the material is transported out.  The combination of the drag time step criterion and spatial resolution requirement mean that these can become prohibitively expensive when the drag is large. That is, when ${\rm St} < 1$, which typically occurs for small dust grains or dense gas regions. 

Note that it is possible to overcome some of the strict time and spatial requirements using implicit methods. For example, \citet{lab14, lab15} used an analytic expression for the relative velocity between gas and dust particles, providing significant efficiency gains when the stopping time is small. \citet{stoyanovskaya-etal18} side stepped the spatial resolution requirement by calculating the drag force using spatial cells, rather than via direct particle-to-particle interaction.

\subsection{The one-fluid method in Phantom}

In the one-fluid method, only one set of particles is used to represent the mixture of gas and dust. Discretizing Equations~(\ref{d-rho2}), (\ref{dv2}), and (\ref{du2}) is straightforward as there is only one type of particle with no drag terms. 
Equation~(\ref{d-rho2}) for the total density becomes
\begin{equation}
    \rho_a = \sum_{b} m_b W_{ab}(h_a);~~~~~~~~h_a = 1.2 \left( \frac{m_a}{\rho_a}\right)^{1/3}. 
	\label{sph-rho-g}
\end{equation}
Equation~(\ref{dv2}) is discretized as
\begin{equation}
\begin{aligned}
\frac{{\rm d}\mathbf{v}_a}{{\rm d}t} =
    & -\sum_b m_b \Bigg [ \frac{P_{{\rm g},a}}{\Omega_a \rho_a^2}\nabla_a W_{ab}(h_a)\\ 
    & \hspace{1cm} + \frac{P_{{\rm g},b}}{\Omega_b \rho_b^2}\nabla_a W_{ab}(h_b)\Bigg ], 
	\label{dva/dt-p} 
\end{aligned}
\end{equation}
and Equation~(\ref{du2}) becomes
\begin{equation}
    \frac{{\rm d}u_a}{{\rm d}t} = 
       \frac{P_{{\rm g},a}}{\rho_a^2 \Omega_a}\sum_b m_b \mathbf{v}_{ab}\cdot\nabla_a W_{ab}(h_a). 
	\label{dua/dt-p}
\end{equation}

Discretizing Equation~(\ref{d-epsilon}) does not place any constraint that the dust fraction should be between $0 \leq \epsilon \leq 1$. Mapping the dust fraction to a function whose codomain is only defined from $[0,1]$ can prevent $\epsilon$ from becoming unphysical. \citet{ballabio-etal18} suggest a parameterization that meets this criterion, given by
 \begin{equation}
    \epsilon = \frac{s^2}{1 + s^2} ,
	\label{epsilon-s}
\end{equation}
such that
\begin{equation}
     s=\sqrt{\frac{\epsilon}{1-\epsilon}}.
\end{equation}
In this formulation, $s$ is then related simply to the ratio of dust-to-gas densities, $s = \sqrt{\rho_{\rm d}/\rho_{\rm g}}$. By substituting the time derivative of the new parameter
 \begin{equation}
    \frac{{\rm d}s}{{\rm d}t} = \frac{1}{2s(1-\epsilon)^2}\frac{{\rm d}\epsilon}{{\rm d}t} ,
	\label{ds}
\end{equation}
in Equation~(\ref{d-epsilon}), the following can be obtained:
 \begin{equation}
    \frac{{\rm d}s}{{\rm d}t} = -\frac{1}{2\rho(1-\epsilon)^2} \Bigg \{ \nabla \cdot \Big[s(1-\epsilon)t_{\rm s}\nabla P_{\rm g}\Big] + (1-\epsilon) t_{\rm s} \nabla P_{\rm g} \cdot \nabla s\Bigg \}.
	\label{ds/dt}
\end{equation}
The SPH discretization is implemented in the form
\begin{equation}
    \frac{{\rm d}s_a}{{\rm d}t} = -\frac{1}{2\rho_a(1-\epsilon_a)^2}\sum_b \Bigg[\frac{m_b s_b}{\rho_b}(D_a + D_b)(P_a - P_b)\frac{\Bar{F}_{ab}}{|r_{ab}|}\Bigg] ,
\end{equation}
where $D_a =  t_{{\rm s},a}(1-\epsilon_a)$, $\Bar{F}_{ab} = [F_{ab}(h_a) + F_{ab}(h_b)]/2$, and $F_{ab} \equiv (\sigma / h^4) f'(q)$.
This equation conserves linear and angular momentum, energy, and mass -- at least up to the accuracy of the time stepping algorithm. 

In addition, the diffusion equation for the dust fraction (Equation~\ref{d-epsilon}), which models the distribution of dust in the gas fluid, imposes a constraint on the time step in addition to the Courant time step \citep{pl15}. This time step criterion is given by 
\begin{equation}
    \Delta t < \Delta t_{\epsilon} \equiv C_0 \frac{h^2}{\epsilon t_{\rm s} c_{\rm s}^2},
	\label{1fluid dt}
\end{equation}
where $C_0$ is a dimensionless safety factor of order unity. Note that this time step criterion is inversely proportional to the stopping time, which is opposite the time step criterion for the two-fluid method (via Equation~\ref{deltaTdrag}). Thus, this time step criterion becomes limiting as the stopping time becomes large, occurring when dust grains are large or in regions of low gas density. \citet{ballabio-etal18} derive a general form of the time step criterion accounting for gradients in $\epsilon$. As with their recommendation, we use the simpler criterion given by Equation~(\ref{1fluid dt}) since it avoids calculating these $\epsilon$ gradients.

\subsubsection{Limiting the stopping time}
\label{sec:1f-limiter}

As mentioned in Section~(\ref{sec:TVA}), the one-fluid method relies on the terminal velocity approximation. This remains accurate when St $\leq 1$, a condition typically met by small dust grains. However, when St $> 1$, numerical inaccuracies can be introduced.

\citet{ballabio-etal18} suggest limiting the stopping time on particles where the stopping time exceeds the local dynamical time, replacing the stopping time ($t_{\rm s}$) by the limited stopping time ($\Tilde{t}_{\rm s}$), given by
\begin{equation}
    \Tilde{t}_{\rm s} = \min (t_{\rm s}, h / c_{\rm s}) .
	\label{limitedTs}
\end{equation}
This restricts changes in the dust fraction for particles that do not adhere to the terminal velocity approximation, curbing potential errors in dust mass conservation. 

Another way to view the limiter is that there is a spatial resolution requirement for the one-fluid method that is analogous to the resolution requirement for the two-fluid method (Equation~\ref{eq:2f-resolution}). That is
\begin{equation}
    h \gtrsim c_{\rm s} t_{\rm s} .
\end{equation}
The smoothing length must be larger than the coupling scale of gas and dust. For the two-fluid method, the resolution must be sufficient to resolve the length scale over which gas and dust would begin comoving. For the one-fluid method, the terminal velocity approximation assumes that the gas and dust are comoving below the resolution scale. Violation of this assumption is problematic.

\subsection{Numerical Stokes numbers in SPH}
\label{sec:stokes-sph}

The Stokes number is defined as the ratio of stopping time to dynamical time; that is, St = $t_{\rm s} / t_{\rm dyn}$ as given by Equation~\ref{eq:stokes}. It is a dimensionless quantity that measures the degree of coupling between the gas and dust. It would seem sensible for turbulence simulations to use the characteristic timescale of the turbulence as the dynamical time (see Section~\ref{sec:initcond}).

However, we find that the results of our simulations are understood best in the context of a numerical Stokes number defined in terms of the local dynamical time, $t_{\rm dyn} = h / c_{\rm s}$. This yields St$_{\rm num} = t_{\rm s} / (h / c_{\rm s})$, where $t_{\rm s}$ is the stopping time. This means the Stokes number is resolution dependent. This local, numerical Stokes number St$_{\rm num}$ is more important in terms of understanding the numerics of the algorithms, as will be shown.

Defining the numerical Stokes number in terms of the local dynamical time introduces a resolution dependence because the dynamical time is proportional to $h$. This means that the numerical Stokes number increases as $h$ decreases, even if the density and dust grain properties are held constant. Therefore, even if the terminal velocity approximation in the one-fluid method is valid at one resolution, there is always a larger resolution for which this will no longer be true (see discussion on the one-fluid limiter in Section~\ref{sec:1f-limiter}).

For the two-fluid method, the problem is reversed. The numerical Stokes number goes down as $h$ increases (i.e., decreasing resolution). This means that there is a point where the spatial resolution requirement is no longer satisfied via Equation~(\ref{eq:2f-resolution}).

Calculating the numerical Stokes number requires the dust stopping time. For the one-fluid method, this is calculated using Equation~\ref{stopping-time2}. Each particle carries the total density (gas plus dust density), so it is trivial to calculate. Note that the one-fluid method does not apply the supersonic correction given by Equation~\ref{kwok}. 

The numerical Stokes number for the two-fluid method is less straightforward because the stopping time is calculated per gas-dust particle pair (Equation~\ref{eq:2f-sph-ts}). Each dust particle has a stopping time with each gas particle it interacts with. We desire a single Stokes number per dust particle. To estimate a single stopping time for dust particles, the gas density is interpolated at the location of the dust particle, and the combined dust plus interpolated gas density is used to calculate a single stopping time per dust particle via Equation~(\ref{stopping-time2}). The gas velocity is also interpolated at the location of the dust particle so that the supersonic correction term can be calculated, which requires the differential velocity between the gas and dust.

\section{Simulation conditions}
\label{sec:initcond}

% Commercon: 1 g/cm^3 grain density, 4.4e-21 g/cm^3 total density. Mach 10
% roughly 2/3'rds our Stokes number

% Hopkins: 2.4 g/cm^3 grain density, 3.82e-23 g/cm^3 total density, Mach 10, assuming mean molecular weight of 2.3 (atomic units?). 
% via their equation 4, an alpha=0.1 equates to 3um in our conditions.

We conduct experiments of driven, isothermal, supersonic turbulence in a periodic box. Simulations are performed using the one- and two-fluid methods for dust grains of size 3 $\mu$m and 10 $\mu$m. The intrinsic density of dust grains is 3 g cm$^{-3}$, representing a combination of carbonaceous (2.2 g cm$^{-3}$) and silicate grains (3.5 g cm$^{-3}$). The box size is $x$, $y$, $z$ $\in$ [0, $L$] with $L = 3$ pc per side. The initial density is uniform with a mean total density (gas plus dust) of $\rho_0 = 10^{-20}$g cm$^{-3}$. This corresponds to a number density of $n \sim 2.6 \times 10^3$ cm$^{-3}$ for a mean molecular weight of 2.3. We use an isothermal equation of state ($\gamma = 1$) with a sound speed $c_{\rm s}$ = 0.2 km s$^{-1}$, corresponding to a temperature of $T \approx$ 11.5 K. The equation of state is well represented by an isothermal equation of state for these densities \citep{scalo-etal98, bt07}. Simulations are performed using $256^3$ gas and $256^3$ dust particles for our two-fluid calculations, and $256^3$ mixed gas-dust particles for our one-fluid calculations.

The turbulence is driven to keep an rms velocity of Mach 10 ($\mathcal{M} = 10$). The simulation is performed for 12 turbulent crossing times, defined as $t_{\rm c} = L/2\mathcal{M} c_{\rm s} \approx$ 0.733 Myr. The driving pattern uses a force derived from an Ornstein-Uhlehnbeck process \citep{ep88, schmidt-etal09, federrath-etal10}. The autocorrelation timescale of the driving is 1 $t_{\rm c}$, meaning that statistically independent snapshots are produced every turbulent crossover time. The driving is purely solenoidal. All calculations utilize the same driving pattern.

The initial numerical Stokes number of our turbulence inside a periodic box setup can be obtained through the Stokes number definition, St = $t_{\rm s} / t_{\rm dyn}$ (Equation~\ref{eq:stokes}), using $t_{\rm dyn} = h / c_{\rm s}$. The stopping time is calculated using Equation~(\ref{stopping-time2}). The initial conditions have zero velocity, meaning the supersonic correction term on the stopping time can be ignored. Additionally, the (initially uniform) smoothing length, $h$, can be recast via Equation~(\ref{h}) in terms of the number of particles in our box, $n$, using the particle mass given by $m_a = \rho_0 L^3 / n$. Recasting this in terms of typical dust grain sizes and conditions for molecular clouds yields 
\begin{equation}
\begin{aligned}
    {\rm St}_{{\rm num},a} = 0.04332&\left (\frac{\rho_{\rm grain}}{3 \, \text{g cm}^{-3}} \right )\left (\frac{s_{\rm grain}}{0.1 \, \mu \text{m}} \right ) \left (\frac{n}{256^3} \right )^{1/3}\\
     &\times \left (\frac{\rho_a}{10^{-20} \, \text{g cm}^{-3}} \right )^{-1} \left (\frac{L}{3 \, {\rm pc}} \right )^{-1}.
\end{aligned}
\label{eq:initial stokes}
\end{equation}
The conditions in Equation~(\ref{eq:initial stokes}) reflect typical dust grain sizes and conditions for molecular clouds. Importantly, the numerical Stokes number increases as the resolution of the calculation increases. This is in line with expectations. Note that this equation assumes that there is no initial velocity, and that the initial density is uniform, as is the case for our initial conditions. The Stokes number will change per particle as turbulence develops.

Via Equation~(\ref{eq:initial stokes}), dust grains with sizes 3 and 10 $\mu$m in these initial conditions correspond to initial Stokes numbers of St$_{\rm num} = 1.3$ and St$_{\rm num}= 4.3$, respectively. These Stokes number regimes are specifically chosen to avoid excessively high resolution requirements for the two-fluid method (via Equation~\ref{eq:2f-resolution}). Our goal is to probe to the numerical accuracy of the two dust solvers in the St$_{\rm num}\sim1$ regime, as this represents the boundary at which numerical issues may possibly manifest. For density variations of 100$\times$ increase and decrease around the mean density, which can be expected for isothermal, supersonic turbulence, this implies Stokes number variations of approximately 20$\times$. For the one-fluid method, it is thus expected that the terminal velocity approximation will be reasonably upheld for a significant fraction of particles.

\section{Results}
\label{sec:results}

Our analysis focuses on the time-averaged statistical properties of turbulent dust and gas. The initial turbulent crossing from $t/t_{\rm c}=0$ to $t/t_{\rm c}=2$ is excluded from our analysis as the turbulence is not fully developed during this phase. We perform averaging over 11 snapshots in the range $t/t_{\rm c}= 2$--12, each taken 1 $t_{\rm c}$ apart such that they are statistically independent snapshots. This time period is chosen to ensure that the averaging is only done once the turbulent cascade has developed.

\subsection{Gas density}
\label{sec:gas_density}

\begin{figure*}
  \centering
  \includegraphics[width=\textwidth]{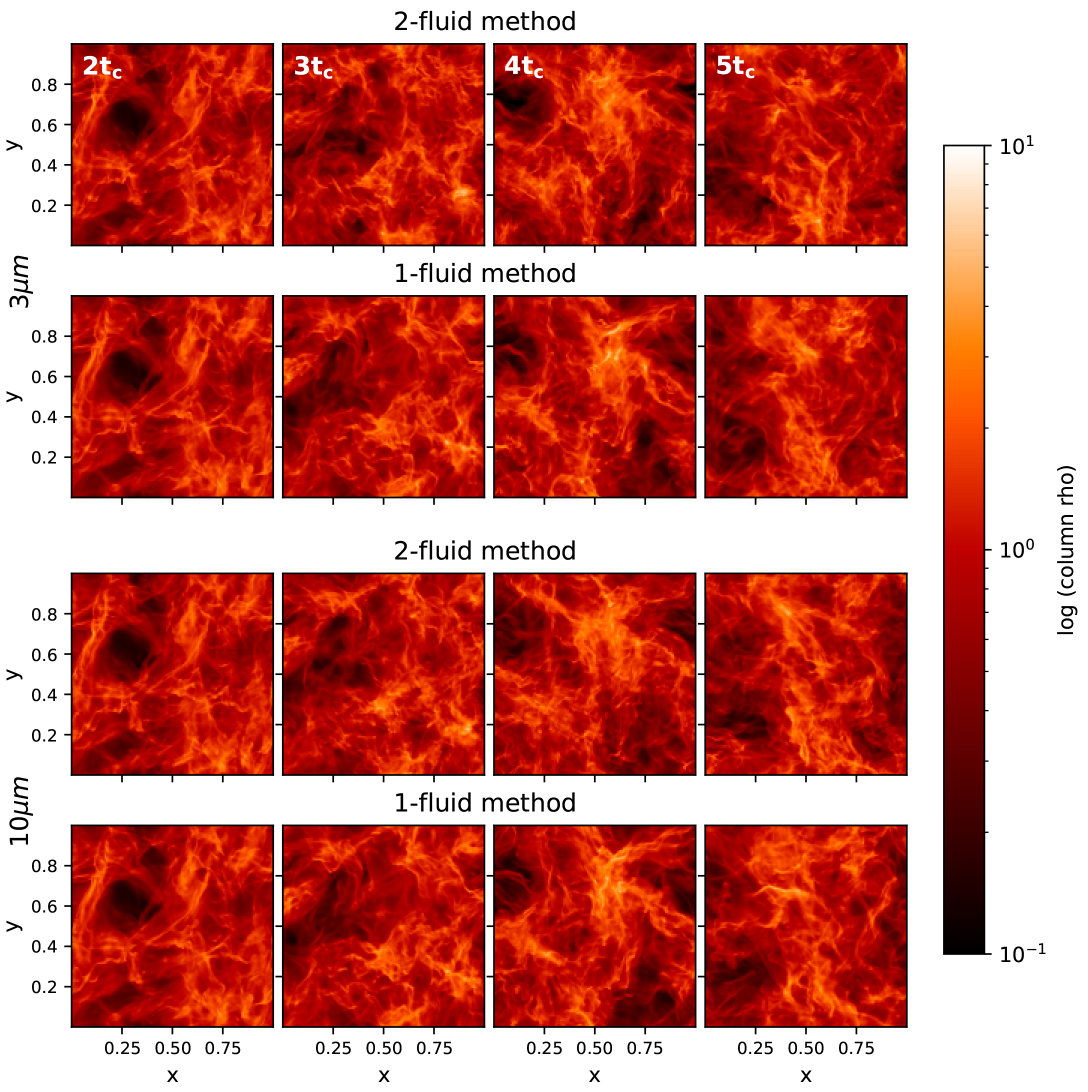} 
  \caption{Column density of the gas with 3 $\mu$m dust grains (top panel) and 10 $\mu$m dust grains (bottom panel) in the two-fluid method (top row of each panel) and in the one-fluid method (bottom row of each panel) at $t/t_{\rm c}$ = 2, 3, 4, and 5 from left to right columns, respectively.}
  \label{fig:gas-40-100}
\end{figure*}

Figure~(\ref{fig:gas-40-100}) shows column-integrated gas densities at $t/t_{\rm c}$ = 2, 3, 4, 5 for 3 $\mu$m and 10 $\mu$m dust grains using the two- and one-fluid methods. At $t / t_{\rm c} = 2$, the gas column densities in both methods are comparable. However, they diverge over time. While they still show the same large-scale structure, different structures appear on smaller scales. Though all calculations use the same driving pattern, the chaotic nature of the turbulence means that individual realizations of the density field will be different across each calculation.

\begin{figure}%
    \centering
    {\includegraphics[width=0.9\columnwidth]{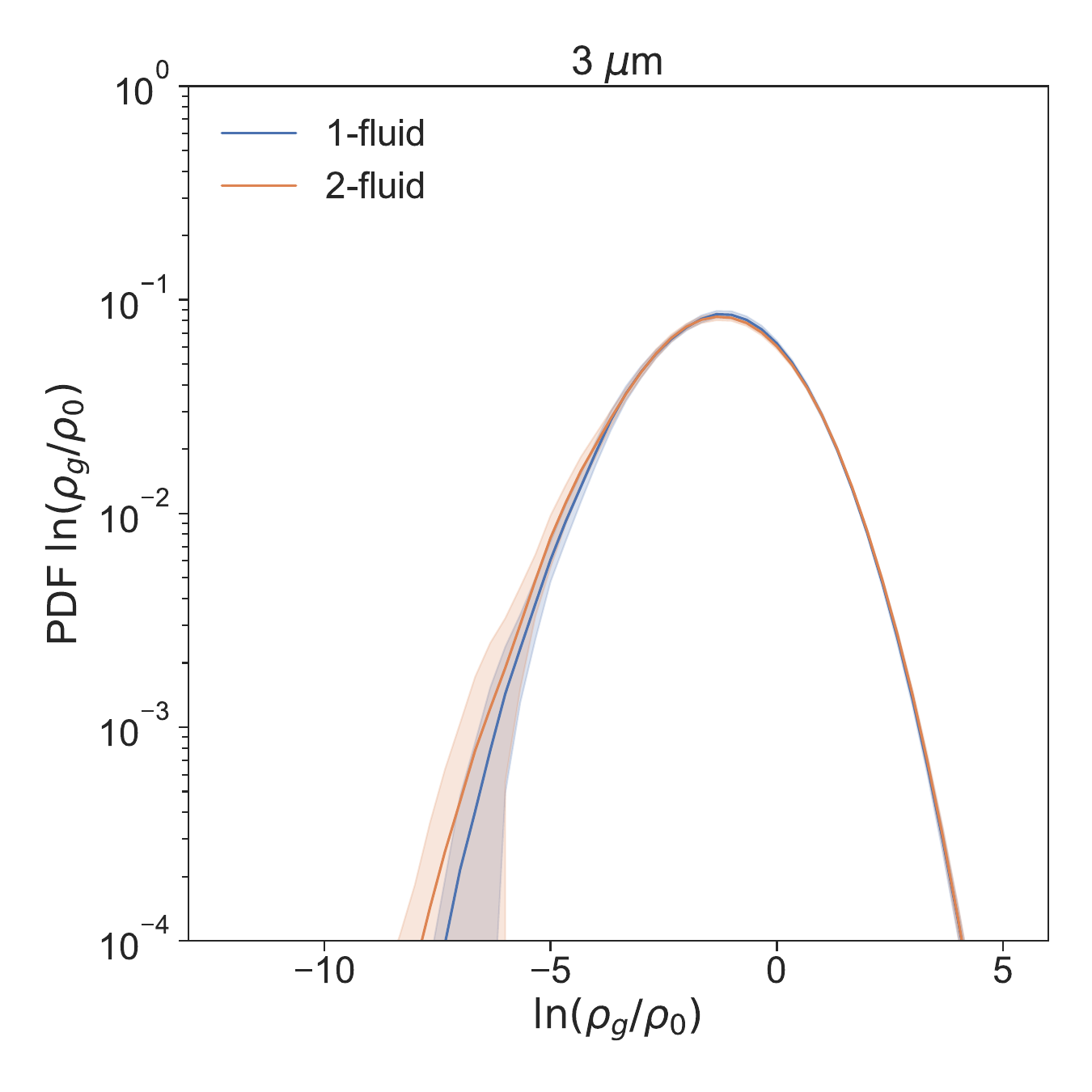} } \\
    {\includegraphics[width=0.9\columnwidth]{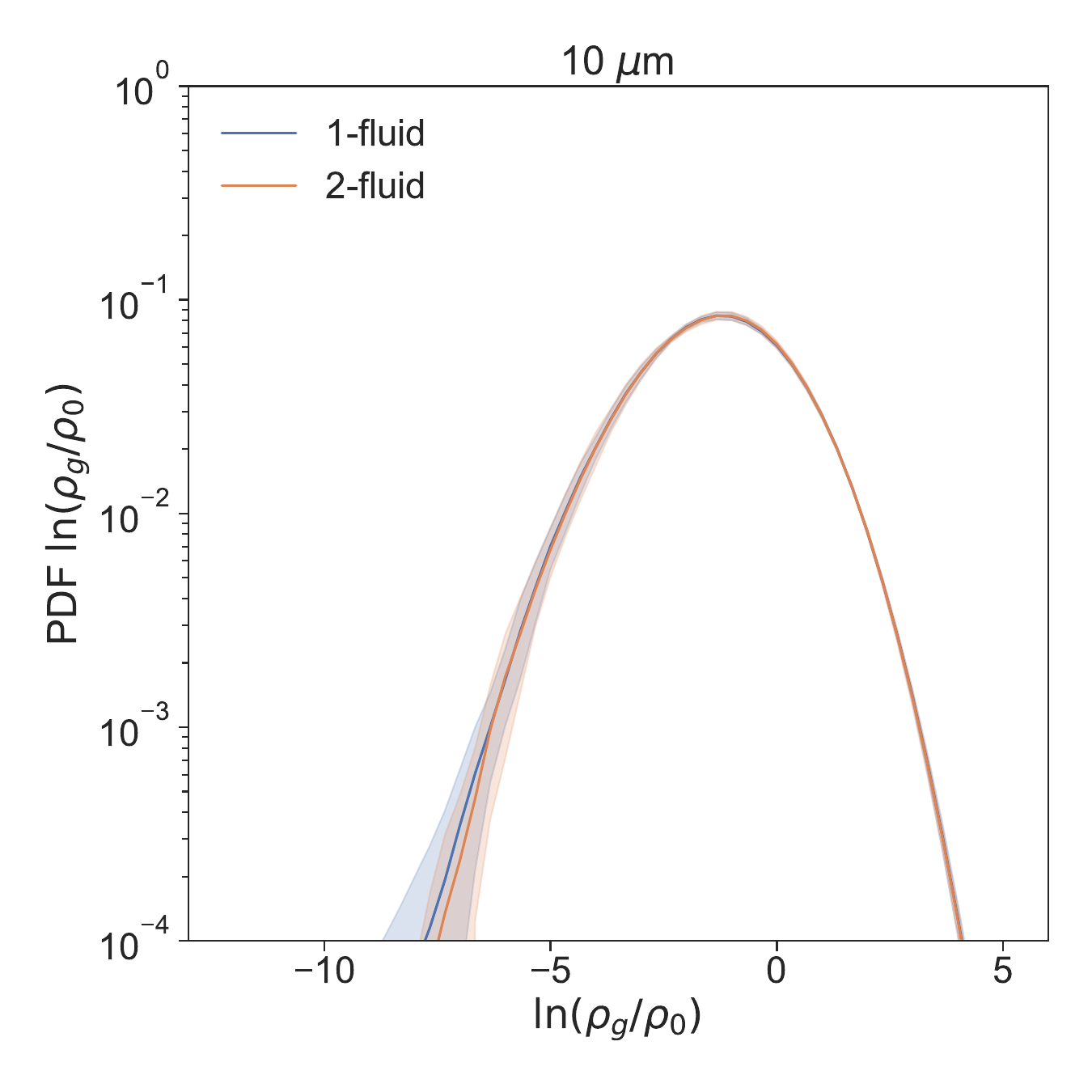} }%
    \caption{Time-averaged volume-weighted PDFs of ln$(\rho_{\rm g}/\rho_0)$ for 3 $\mu$m (top) and 10 $\mu$m (bottom) dust grains in $256^3$ resolution in one- and two-fluid methods. Shaded regions represent the standard deviation of the time averaging.}
    \label{fig:PDF-gas}%
\end{figure}

Figure~(\ref{fig:PDF-gas}) shows the time-averaged, volume-weighted probability density functions (PDFs) of the gas density for 3 $\mu$m and 10 $\mu$m dust grains. The volume-weighted PDFs are calculated by interpolating the particle densities for each turbulent turnover snapshot onto a grid with a resolution of 512$^3$ cells, i.e., twice the number of particles. The PDFs are not calculated using the particles directly because this would lead to a mass-weighted PDF. Time averaging helps to mitigate the effects of turbulent variability and is necessary to ensure statistically meaningful results. 

The gas PDFs exhibit a lognormal distribution, characteristic of supersonic, isothermal turbulence \citep[e.g.,][]{vs94, pvs98, federrath-etal10}. They can be fit in log-space to a normal distribution with Table~(\ref{tbl:gasPDFs}) giving the mean and standard deviation of each fit. No significant difference in the gas PDFs are found between the simulations with 3 $\mu$m and 10 $\mu$m dust grains, or between the two-fluid and one-fluid methods. 

\begin{deluxetable}{llcc}
\tablecaption{Lognormal fits of the gas density Probability Density Functions\label{tbl:gasPDFs}}
\tablehead{
Size ($s$) & Method & Mean & Std.\ dev.\ ($\sigma$)
}
\startdata
3 $\mu$m & Twofluid & $-1.2972 \pm 0.0055$ & $1.5930 \pm 0.0055$ \\
3 $\mu$m & One fluid & $-1.2638 \pm 0.0051$ & $1.5580 \pm 0.0051$ \\
10 $\mu$m & Two fluid & $-1.2748 \pm 0.0056$ & $1.5750 \pm 0.0056$ \\
10 $\mu$m & One fluid & $-1.2839 \pm 0.0053$ & $1.5779 \pm 0.0053$
\enddata
\end{deluxetable}

\subsection{Dust density}

\begin{figure*}
  \centering
  \includegraphics[width=\textwidth]{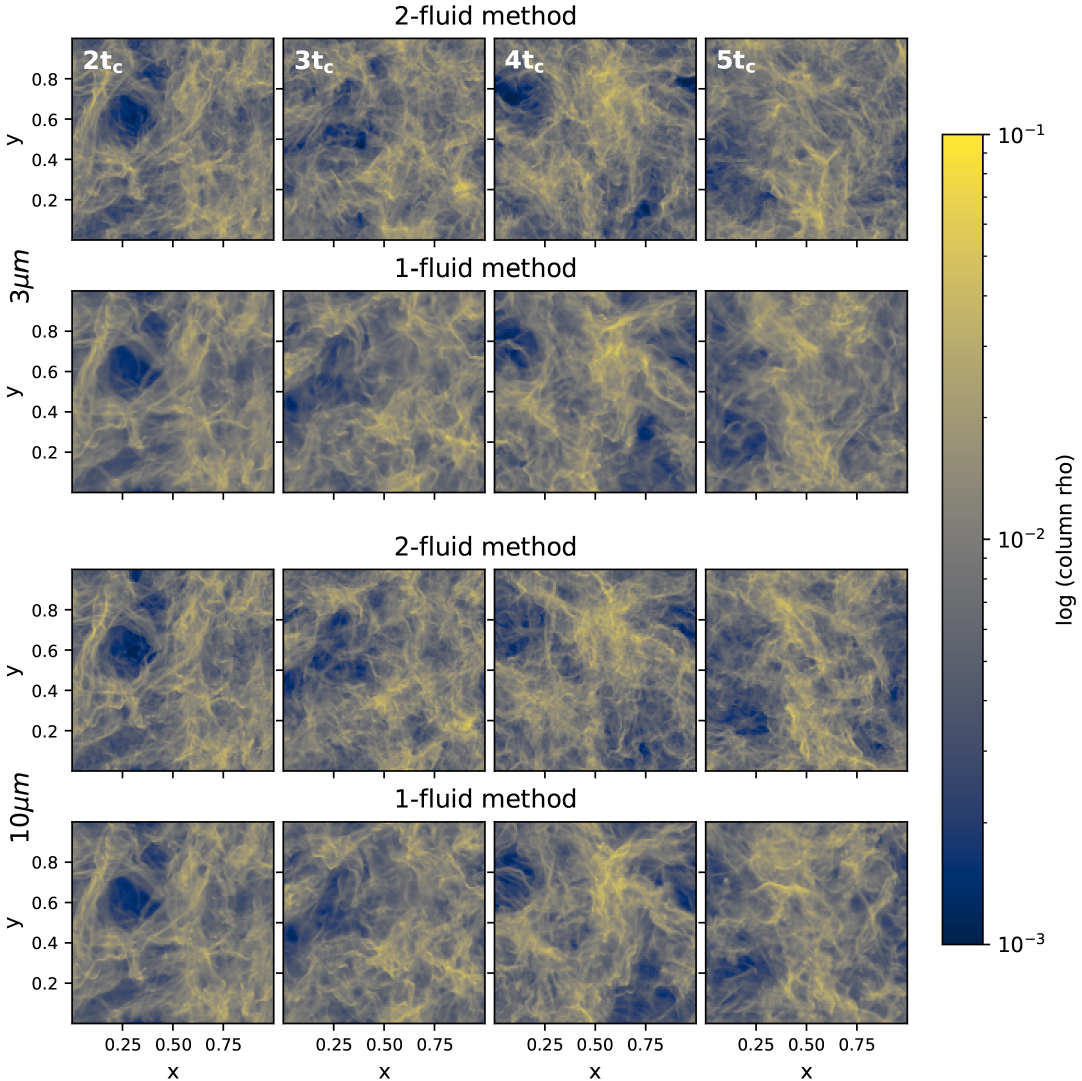} 
  \caption{Column density of the dust in the two-fluid method (top rows) and in the one-fluid method (bottom rows) for 3 and 10 $\mu$m dust grains at $t/t_{\rm c}$ = 2, 3, 4, and 5 from left to right columns, respectively.}
  \label{fig:dust-40-100}
\end{figure*}

Figure~(\ref{fig:dust-40-100}) shows the column-integrated dust density at $t/t_{\rm c}$ = 2, 3, 4, 5 for 3 and 10 $\mu$m dust grains in the two- and one-fluid methods, similar to the gas density renderings in Figure~(\ref{fig:gas-40-100}). The dust density fields are almost exactly the same between the two- and one-fluid results for both grain sizes at $t/t_{\rm c}=2$. The density fields diverge over time. The large-scale structures remain similar at $t/t_c=5$, but small-scale differences are evident. This behavior was also seen in the column-integrated gas density.

\begin{figure}%
    \centering
    {\includegraphics[width=0.9\columnwidth]{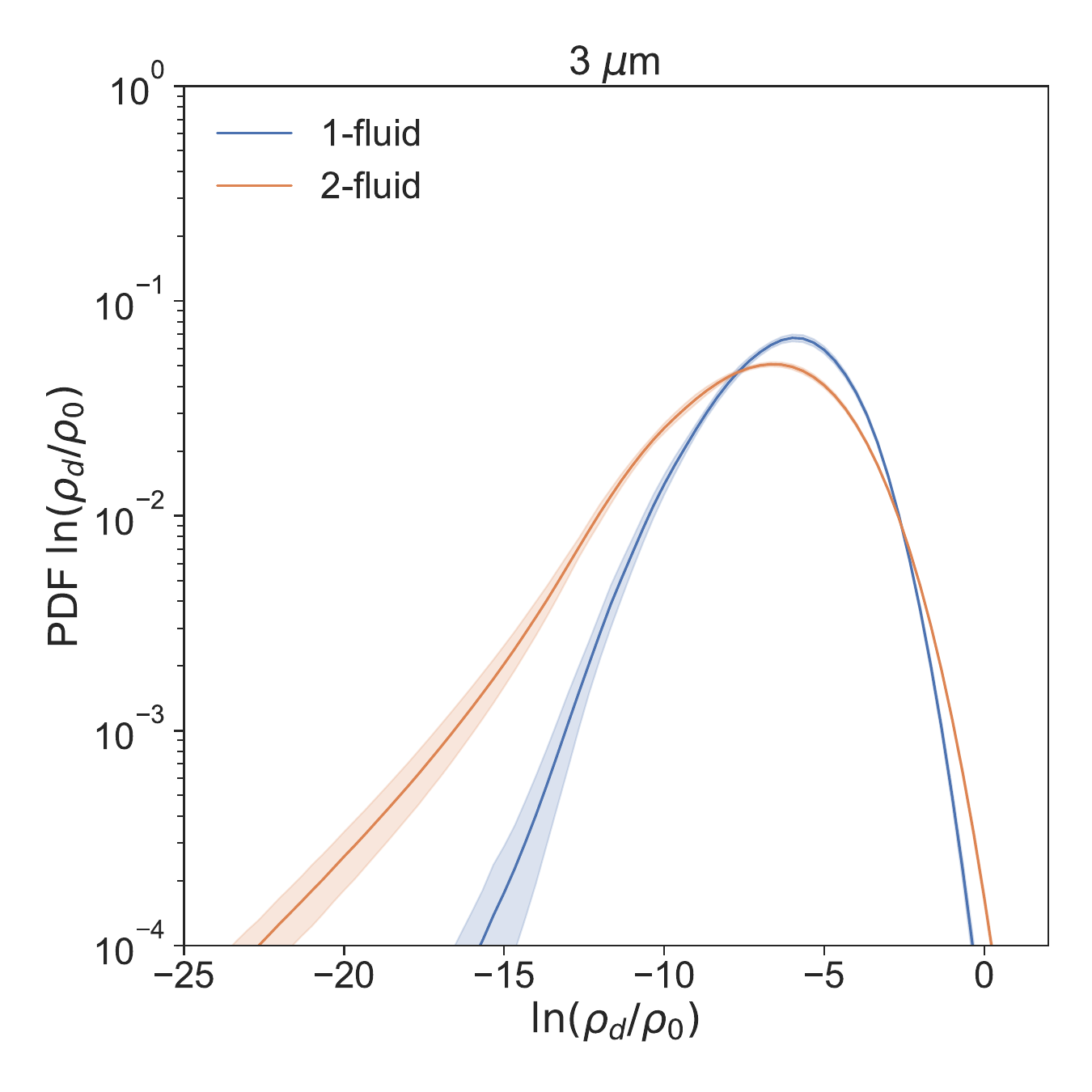} } \\
    {\includegraphics[width=0.9\columnwidth]{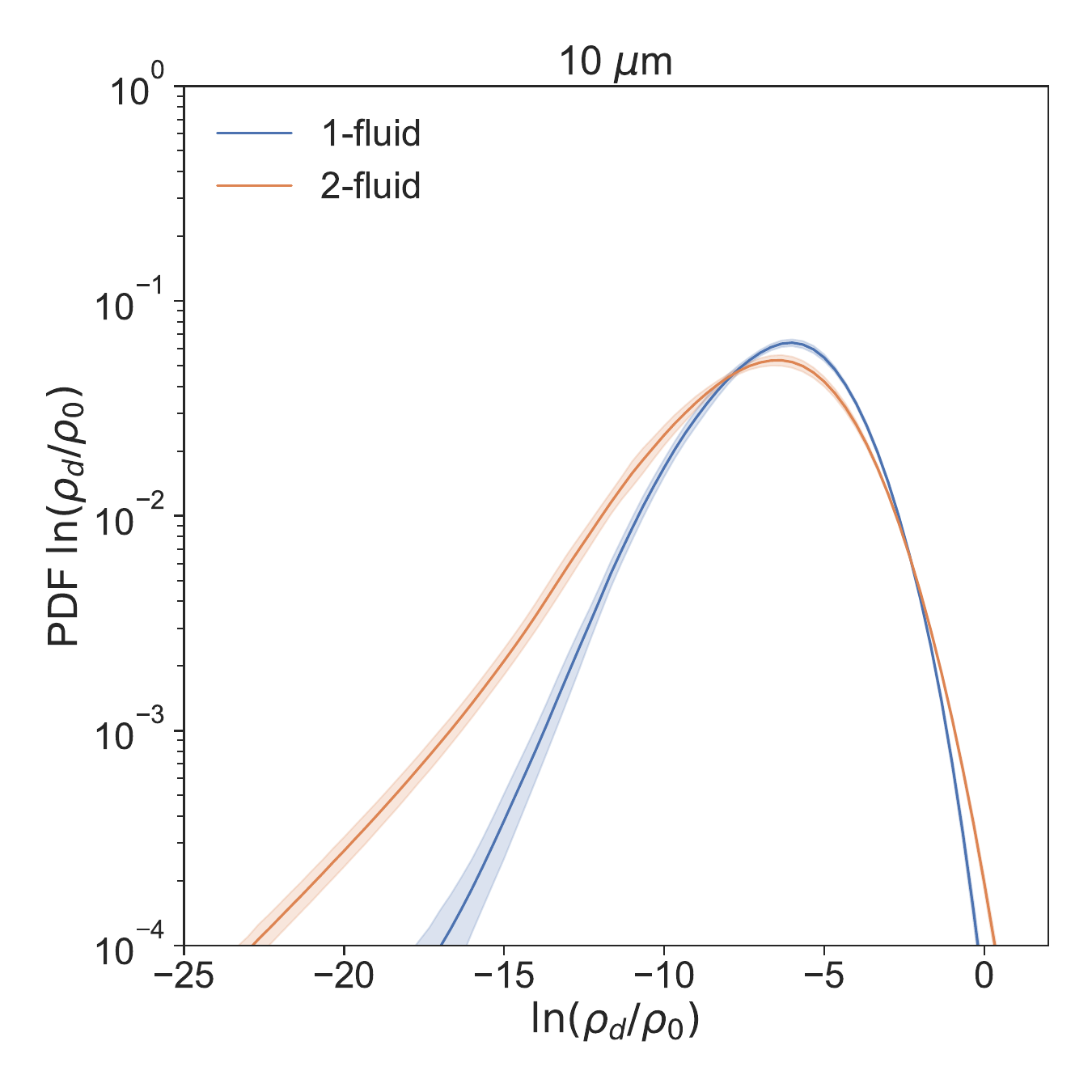} }%
    \caption{Time-averaged volume-weighted PDFs of ln$(\rho_{\rm d}/\rho_0)$ for 3 $\mu$m (top) and 10 $\mu$m (bottom) dust grains in $256^3$ resolution in one- and two-fluid methods. Shaded regions represent the standard deviation of the time averaging.}
    \label{fig:PDF-dust}%
\end{figure}

Figure~(\ref{fig:PDF-dust}) shows the PDFs of dust density for 3 $\mu$m and 10 $\mu$m dust grains. The PDFs for both grain sizes peak at similar values. The peaks are located at lower densities than the gas PDFs, shifted an amount appropriate for the mean dust-to-gas ratio. For both grain sizes, the high-density tails of the dust PDFs are closely aligned for both  methods, though the two-fluid tail extends to slightly higher densities. The largest difference is in the low-end tail. The two-fluid results extend several orders of magnitude lower than the one-fluid results. The implication of this is that the two-fluid dust density PDFs are more asymmetric than the one-fluid PDFs. 

\begin{deluxetable}{llcc}
\tablecaption{Lognormal fits of the dust density Probability Density Functions\label{tbl:dustPDFs}}
\tablehead{
Size ($s$) & Method & Mean & Std dev ($\sigma$)
}
\startdata
3 $\mu$m & Two-fluid & $-6.9776 \pm 0.0270$ & $2.5923 \pm 0.0270$ \\
3 $\mu$m & One-fluid & $-6.1704 \pm 0.0190$ & $1.9784 \pm 0.0190$ \\
10 $\mu$m & Two-fluid & $-6.8726 \pm 0.0296$ & $2.4853 \pm 0.0296$ \\
10 $\mu$m & One-fluid & $-6.3531 \pm 0.0205$ & $2.0863 \pm 0.0205$
\enddata
\end{deluxetable}

Table~(\ref{tbl:dustPDFs}) lists the mean and standard deviations for the dust density PDFs fit to lognormal distributions. The fitted parameters are consistent for the same method, although they are different between the one-fluid and two-fluid methods. The mean values are lower for the two-fluid fits compared to the one-fluid fits, along with a wider standard deviation. Overall, the dust density PDFs have larger standard deviations than the gas density PDFs, in addition to larger errors on the fitted parameters. This means that the dust density PDFs deviate in shape from the gas density PDFs, which can also be understood as a consequence of the asymmetries in the dust density PDFs found in Figure~(\ref{fig:PDF-dust}).

\subsection{Correlation of dust and gas}
\label{sec:dtg}

\begin{figure*}%
    \centering
    {\includegraphics[width=8cm]{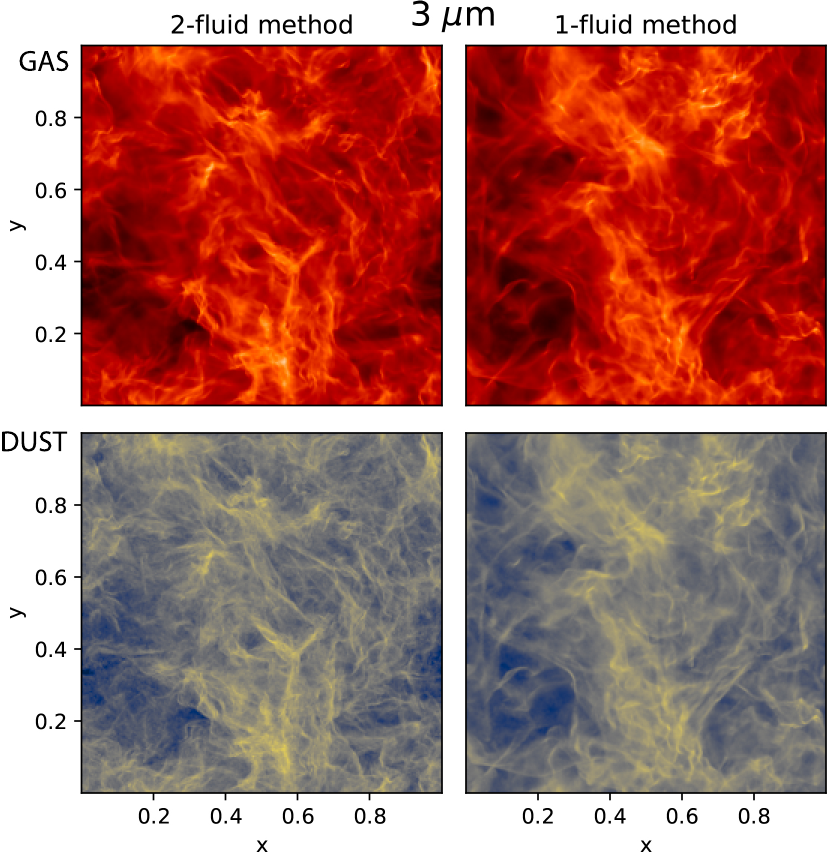} } 
    {\includegraphics[width=8.65cm]{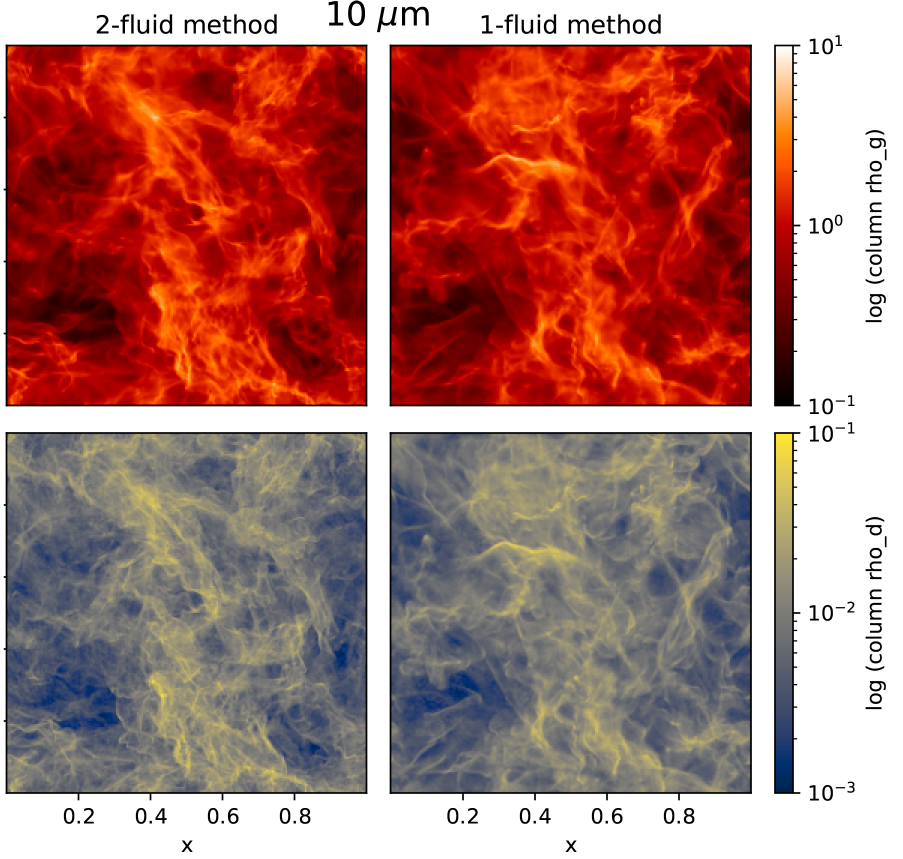} }
    \caption{Column density of the gas (top row) and the dust (bottom row) for 3 and 10 $\mu$m dust grains (left and right panels, respectively) in the two-fluid method (left column of each panel) and the one-fluid method (right column of each panel) at $t/t_{\rm c} = 5 ~(\approx$ 3.67 Myr).}
  \label{fig:t5}
\end{figure*}

Figure~(\ref{fig:t5}) compares the column densities of the gas (top row) and dust (bottom row) for 3 $\mu$m (left panel) and 10 $\mu$m (right panel) dust grains in the two-fluid method (left column of each panel) and the one-fluid method (right column of each panel) at $t/t_{\rm c} = 5$. The large-scale structure of the dust follows the gas in both methods for both grain sizes. However, the dust column density in the two-fluid method appears to exhibit sharper density contrasts on the small scale compared to the gas density. The one-fluid dust density also has sharper density contrasts than the gas, but not as pronounced as the two-fluid dust density.

\begin{figure*}%
    \centering
    {\includegraphics[height=8cm]{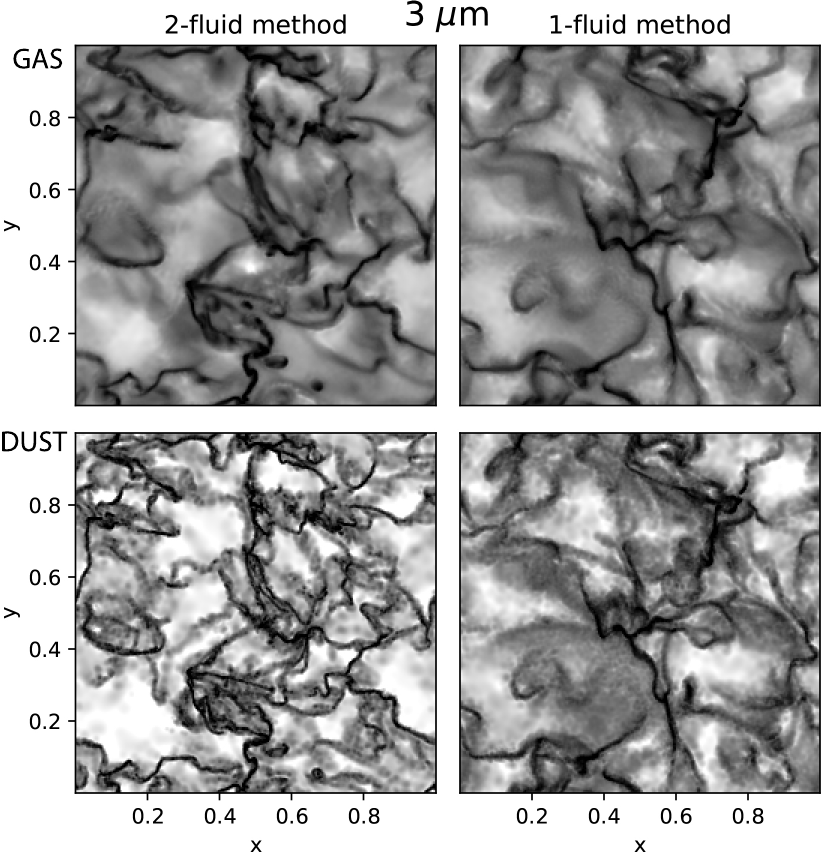} }
    {\includegraphics[height=8cm]{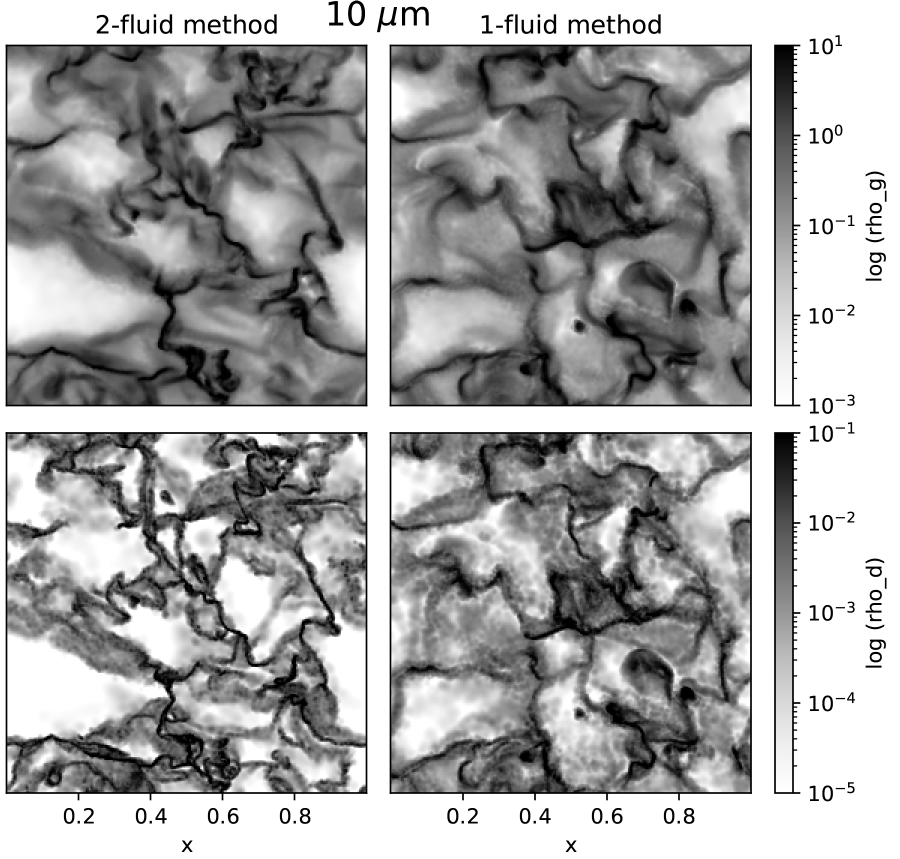} }
    \caption{Slices of the gas (top row) and the dust (bottom row) densities for 3 and 10 $\mu$m dust grains (left and right panels, respectively) in the two-fluid method (left column of each panel) and the one-fluid method (right column of each panel) at $t/t_{\rm c} = 5 ~(\approx 3.67 Myr) $.}
  \label{fig:t5-slice}
\end{figure*}

Figure~(\ref{fig:t5-slice}) shows slices of the dust and gas densities at $t/t_{\rm c}$ = 5 for 3 and 10 $\mu$m dust grains in the two- and one-fluid methods. A similar picture emerges as with the column-integrated densities, but even more pronounced. The large-scale structure of the dust traces the structure of the gas, however the density contrast between low- and high-density regions in the dust are greater than the gas. This appears more pronounced for the two-fluid method than the one-fluid method, suggesting that decoupling occurs more readily for the two-fluid method.

\begin{figure}%
    \centering
    {\includegraphics[width=0.9\columnwidth]{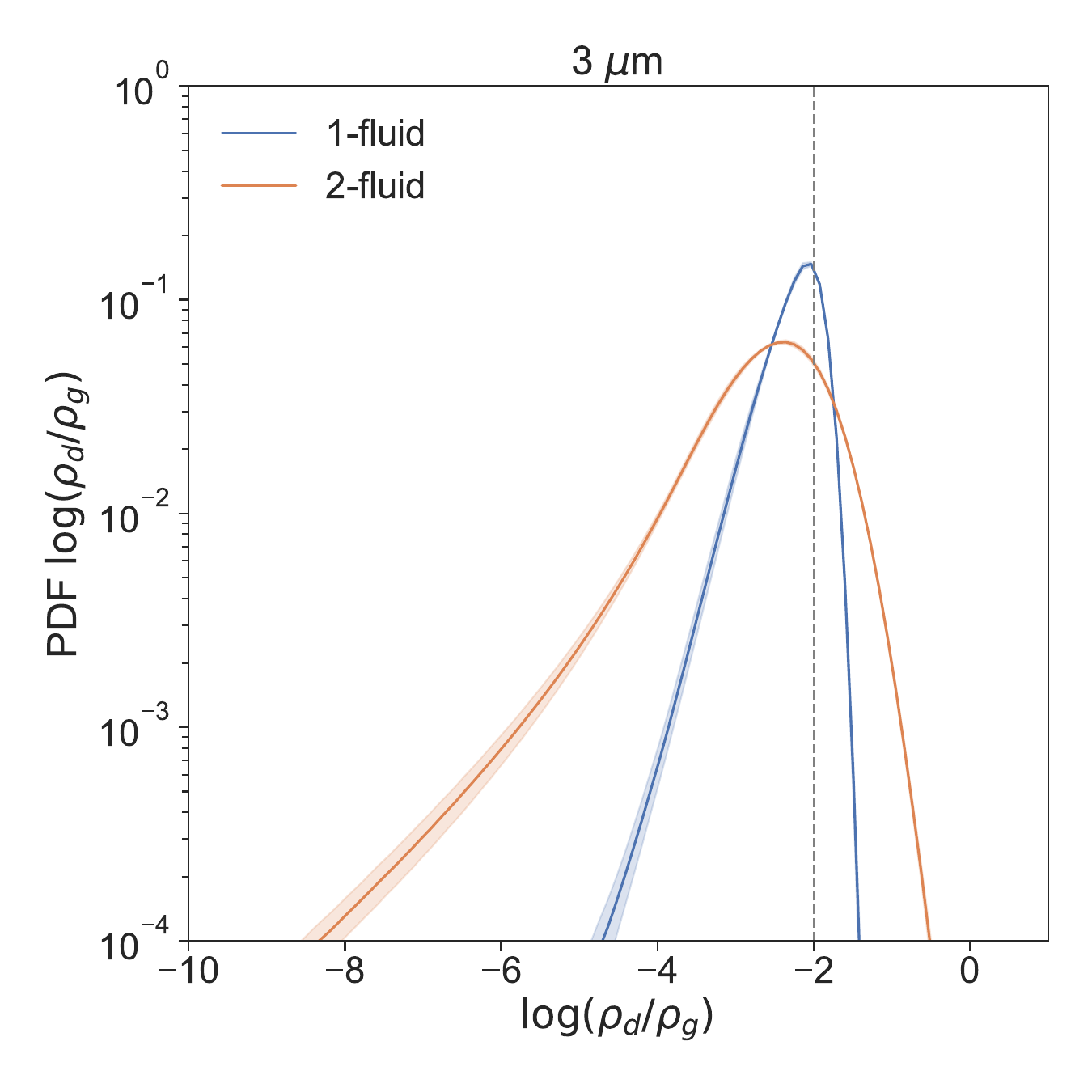} } \\
    {\includegraphics[width=0.9\columnwidth]{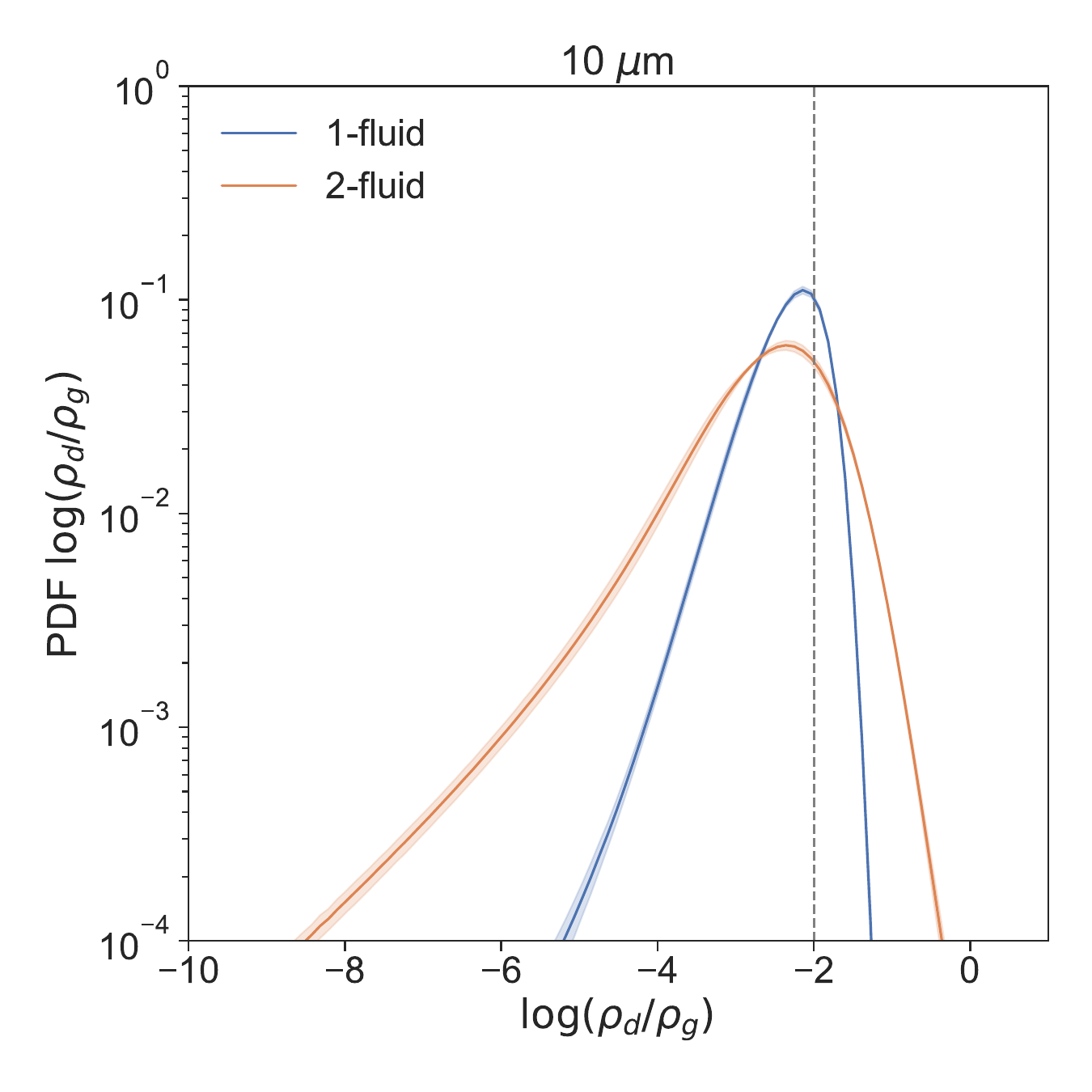} }%
    \caption{Time-averaged volume-weighted PDFs of log$(\rho_{\rm d}/\rho_{\rm g})$ for 3 $\mu$m (top) and 10 $\mu$m (bottom) dust grains in $256^3$ resolution in the two- and one-fluid approaches. Shaded regions represent the standard deviation of the time averaging.}
    \label{fig:dtg_l2f}%
\end{figure}

Figure~(\ref{fig:dtg_l2f}) depicts the PDFs of dust-to-gas ratio for dust grains of sizes 3 and 10 $\mu$m. They both peak at approximately the mean dust-to-gas ratio of 0.01. The PDFs for the two-fluid dust-to-gas ratios are significantly broader than the PDFs obtained using the one-fluid method. Both PDFs are asymmetric, though the two-fluid has a larger skew.

% 99th percentiles
% 0.0196  1f 3mic
% 0.0251  1f 10mic
% 0.0629  2f 3mic
% 0.0681  2f 10mic

The one-fluid method reaches a maximum dust-to-gas ratio of $\sim 0.03$ for 3 $\mu$m and $\sim 0.05$ for 10 $\mu$m dust grains, whereas the two-fluid method reaches a maximum dust-to-gas ratio of $\sim 0.30$ for 3 $\mu$m and $\sim 0.39$ for 10 $\mu$m dust grains, where we quote the maximums as the 99th percentiles of the time-averaged PDFs. This represents a local increase from the mean dust-to-gas ratio of $3$--$5\times$ with the one-fluid method, but a $30$--$39\times$ increase with the two-fluid method. Due to conservation of mass, this increase necessitates a corresponding decrease in the minimum dust-to-gas ratios observed in the results. The minimum dust-to-gas ratios in the PDFs are not significantly different between 3 and 10 $\mu$m dust grains for the two-fluid method, though the PDFs for the one-fluid method are slightly broader for 10 $\mu$m grains compared to 3 $\mu$m grains.

That the dust-to-gas ratio distributions are narrower for the one-fluid results than the two-fluid results can be understood in context of the stopping time limiter used with the one-fluid method. The limiter is applied in low-density regions. As seen in Figure~(\ref{fig:t5-slice}), there is more dust mass in low-density regions for the one-fluid method compared to the two-fluid method. This homogenizes the dust throughout the domain, yielding a narrow distribution of dust-to-gas ratios.

\subsection{Stokes number}
\label{sec:stokes_number}

\begin{figure}%
    \centering
    {\includegraphics[width=0.9\columnwidth]{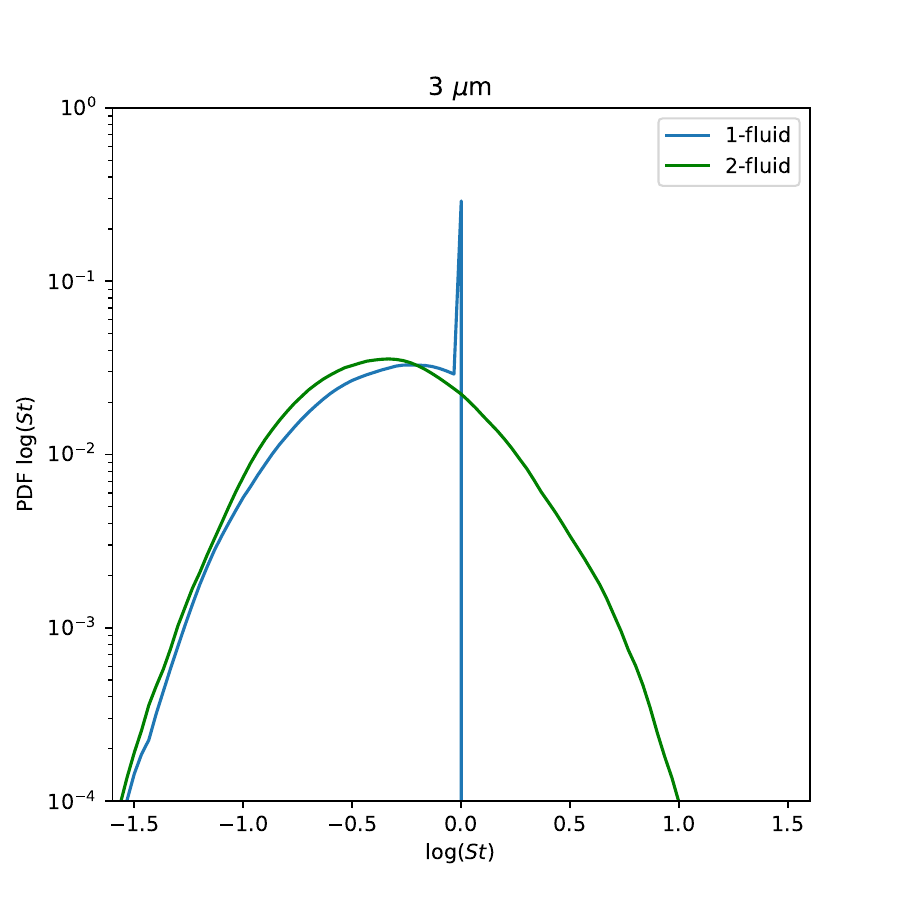} } \\
    {\includegraphics[width=0.9\columnwidth]{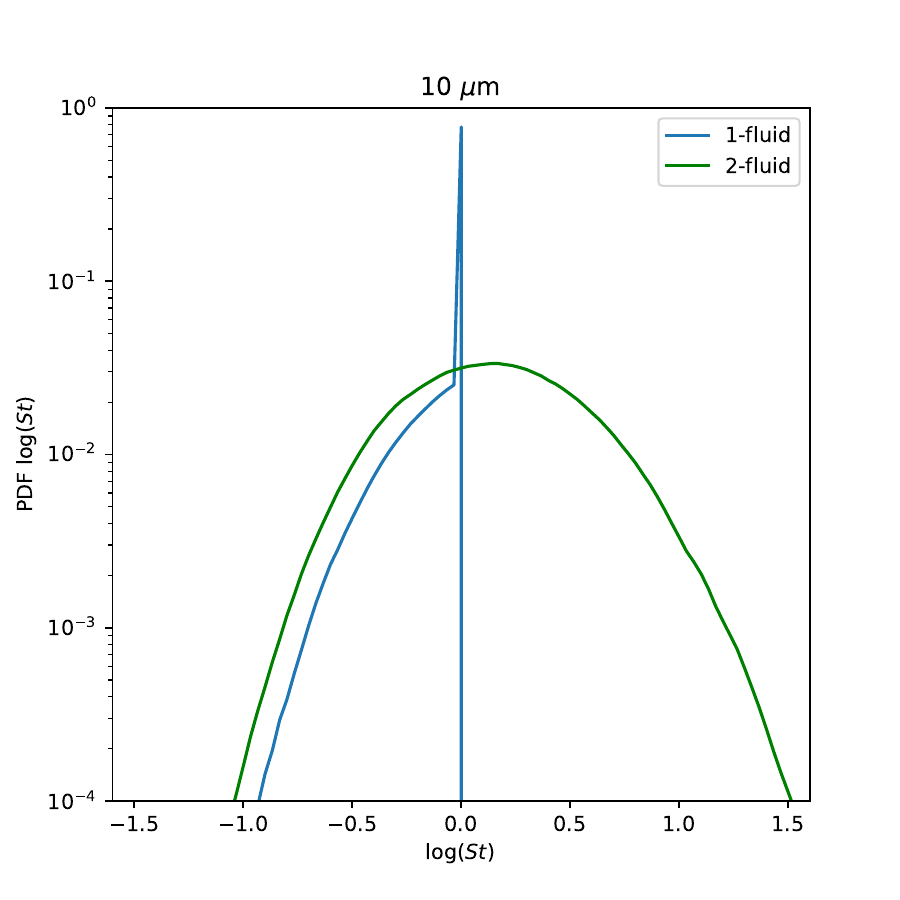} }%
    \caption{PDFs of $\log($St$_{\rm num})$ for 3 $\mu$m (top) and 10 $\mu$m (bottom) dust grains for the two- and one-fluid approaches at $t/t_c$~=~8.}
    \label{fig:stokes}%
\end{figure}
The Stokes number, ${\rm St} = t_{\rm s} / t_{\rm dyn}$ (Equation~\ref{eq:stokes}), is a dimensionless quantity that measures the degree of coupling between the gas and dust. We calculate the numerical Stokes number, St$_{\rm num}$, per particle defined in terms of the local dynamical time, $t_{\rm dyn} = h / c_{\rm s}$, as described in Section~(\ref{sec:stokes-sph}). 

Figure~(\ref{fig:stokes}) shows the probability distributions of the logarithm of Stokes numbers over the SPH particles at $t/t_{\rm c}= 8$. The Stokes number distributions for 3 $\mu$m dust grains have a roughly $3\times$ smaller peak than the peak for the 10 $\mu$m distributions, reflecting the dependence of the Stokes number on grain size (via the stopping time). 

Naively, one would expect the two- and one-fluid Stokes number distributions to be equal since the simulations have the same grain density, grain size, resolution, and gas density distributions (see Section~\ref{sec:gas_density}). This is true below St$_{\rm num} < 1$, with the two- and one-fluid results in good agreement for both the 3 and 10 $\mu$m dust grain calculations. The large spike at St$_{\rm num} = 1$ for the one-fluid results is a consequence of the stopping time limiter (Equation~\ref{limitedTs}). The limiter prohibits stopping times greater than the dynamical time because of the terminal velocity approximation, thereby artificially capping the Stokes number at one. One way to view this is that the dust grains act as though they have smaller size ($s_{\rm grain}$) when they are in low-density regions.

Approximately 30\% and 75\% of the particles are in the St$_{\rm num} > 1$ regime for the 3 and 10 $\mu$m dust grain simulations, respectively. Thus, the one-fluid limiter is applied to approximately 30\% and 75\% of the particles. This results in artificially enhanced coupling. In effect, it is as if the calculations have a population of dust grains ranging from 0.3 to 3 $\mu$m or 0.3 to 10 $\mu$m, reflecting the maximum Stokes number of the 3 and 10 $\mu$m calculations. It is important to note that the limiter is critical for numerical stability in this regime (see Section~\ref{sec:limiter}). 

Conversely, even though the stopping time is not limited for the two-fluid results, the numerical resolution requirement (St$_{\rm num} > 1$, see Equation~\ref{eq:2f-resolution}) implies that this is not upheld for 70\% and 25\% of particles in the 3 and 10 $\mu$m simulations, respectively. This means that the one-fluid results are most accurate for St$_{\rm num} < 1$ and the two-fluid results most accurate for St$_{\rm num} > 1$. 

The distribution of Stokes numbers can be understood from the dependence on density (and indirectly smoothing length), as all other conditions remain constant. In SPH, the mass of each particle is constant. Thus, the smoothing length can be related to density as $h \propto \rho^{-1/3}$ through Equation~(\ref{h}), and the dynamical time expressed in terms of density as $t_{\rm dyn} \propto \rho^{-1/3}$. As the stopping time is proportional to inverse of total density through Equation~(\ref{stopping-time2}), that is, $t_{\rm s} \propto \rho^{-1}$, the numerical Stokes number of a particle therefore scales according to St$_{\rm num} \propto \rho^{-2/3}$.

The expected statistics of the numerical Stokes number distributions can be understood in relation to the gas density PDF. The standard deviation of the two-fluid numerical Stokes number distributions, given in natural logarithm to match the gas density distribution statistics, is 0.87 for 3 $\mu$m and 0.89 for 10 $\mu$m. The gas density PDFs have standard deviations of approximately 1.56 for all simulations (Table~\ref{tbl:gasPDFs}). This is close to the value expected by the St$_{\rm num} \propto \rho^{-2/3}$ scaling relation, keeping in mind that the numerical Stokes number distribution is calculated on the particles and the gas density distribution on the interpolated grid.

\subsection{Impact of the dust limiter}
\label{sec:limiter}

The one-fluid method uses a limiter on the stopping time when St$_{\rm num} > 1$ in order to maintain the terminal velocity approximation. This artificially keeps the dust in the St$_{\rm num} \leq 1$ regime, which in effect causes the dust to be more strongly coupled to the gas. Despite this seemingly undesirable behavior, the stopping time limiter is critical for our calculations to avoid numerical issues, such as nonconservation of dust mass, since approximately 30\% and 75\% of particles are in the St$_{\rm num} > 1$ regime for the 3 and 10 $\mu$m grain size simulations, respectively. 

\begin{figure}%
    \centering
    \includegraphics[width=0.9\columnwidth]{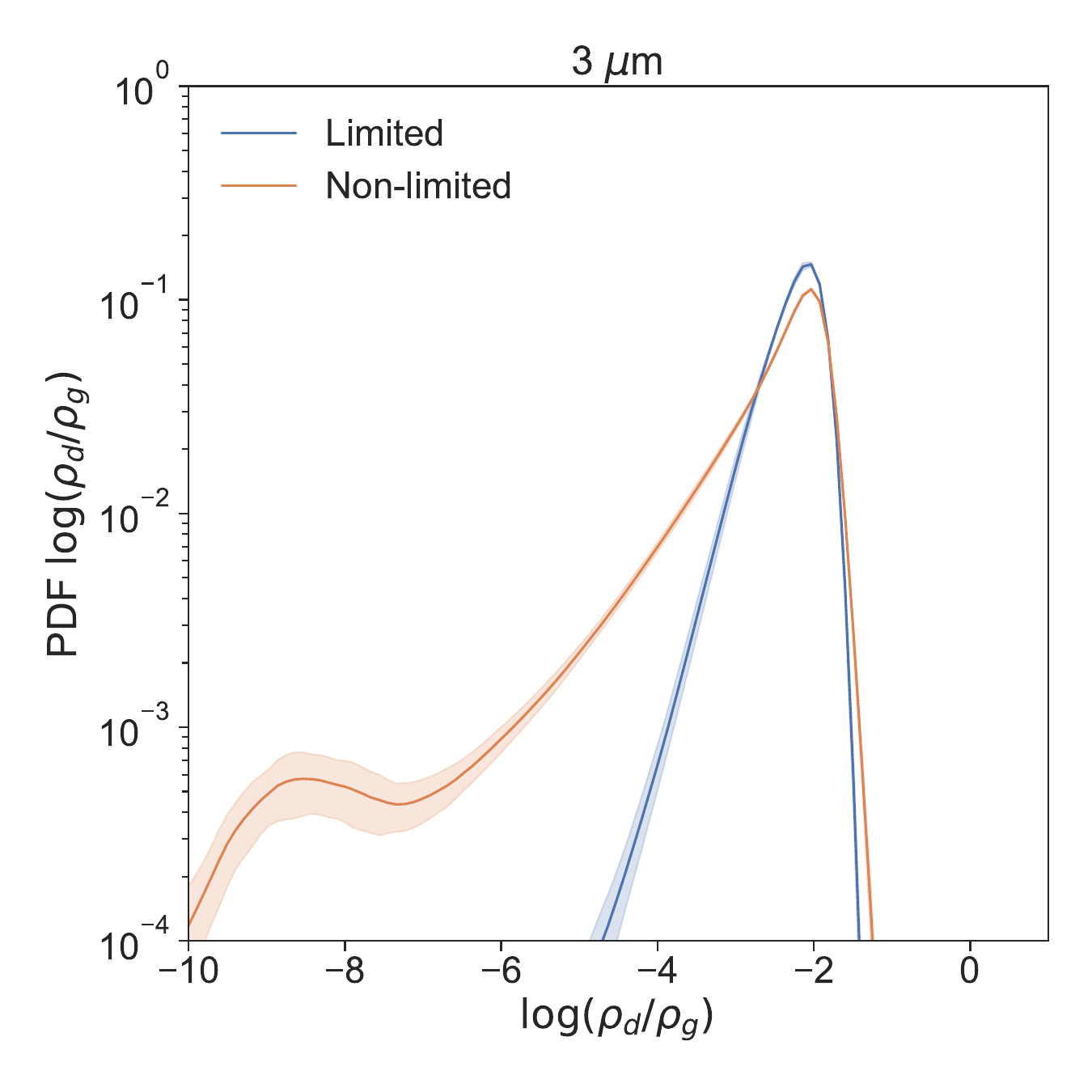} \\
    \includegraphics[width=0.9\columnwidth]{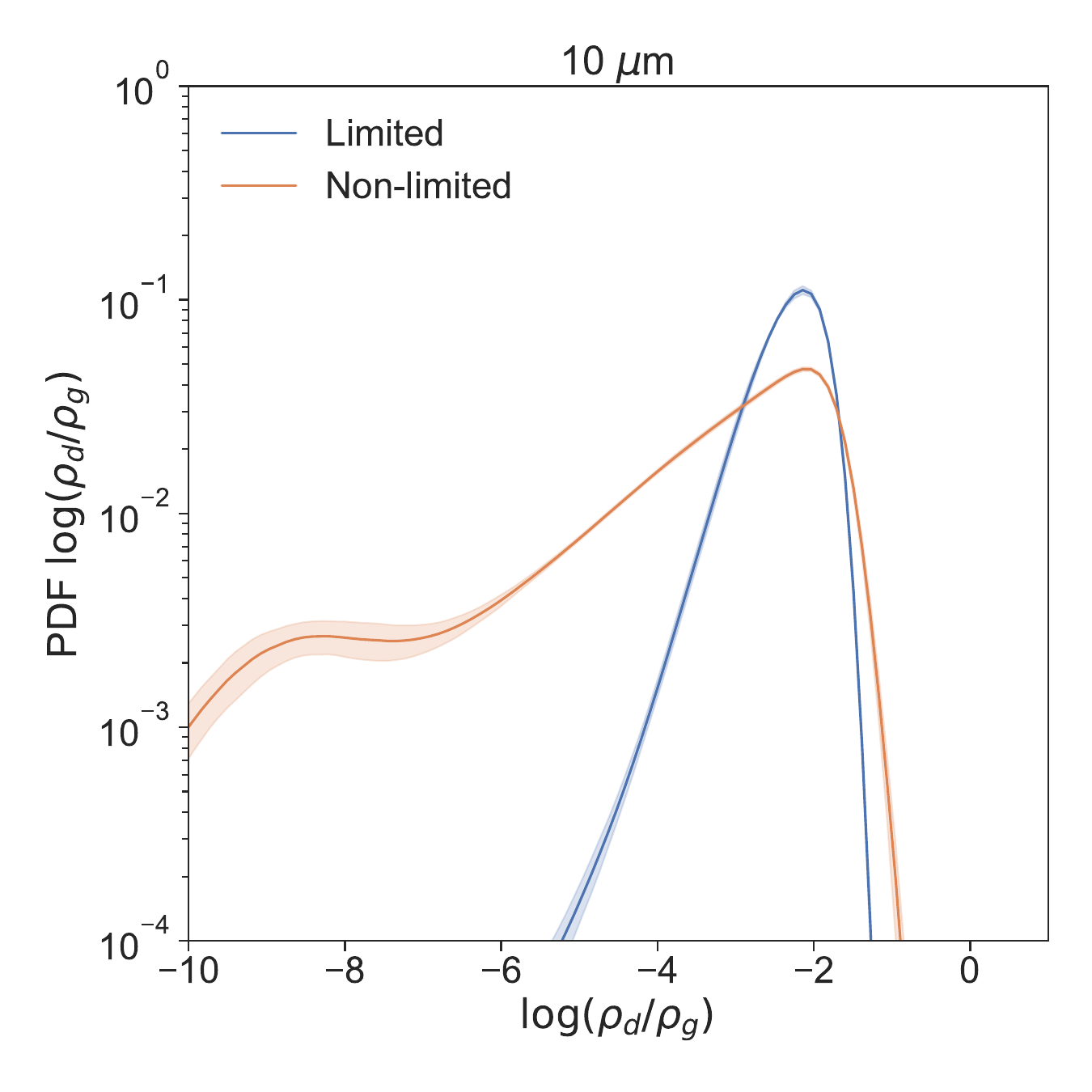} %
    \caption{Time-averaged volume-weighted PDFs of log($\rho_d/\rho_g$)  for 3 $\mu$m (top) and 10$\mu$m (bottom) dust grains in $256^3$ resolution in the limited and nonlimited one-fluid methods. Shaded regions represent the standard deviation of the time averaging.}
    \label{fig:dtg_lnl}%
\end{figure}

Figure~(\ref{fig:dtg_lnl}) shows the PDFs of dust-to-gas ratio for one-fluid simulations with and without the stopping time limiter. Without the limiter, a spurious peak appears at very low dust-to-gas ratios. This is entirely numerical in origin. A similar peak can be found in the results of \citet{tpl17}. This peak disappears when the limiter is used. In fact, the limiter also reduces the range of dust-to-gas ratios, as discussed in Section~(\ref{sec:dtg}).

The right-hand tails of the dust-to-gas ratio PDFs are closely aligned between the limited and nonlimited results. The limiter is only applied to regions where St$_{\rm num} > 1$, implying that if there is agreement in the larger dust-to-gas ratio regions, then this condition occurs most predominantly inside of dense filaments. The right-hand tails for the limited and nonlimited results are in particularly close agreement for the 3 $\mu$m simulations, as a greater fraction (approximately 70\%) of the particles stay below St$_{\rm num} = 1$.

\begin{figure}%
    \centering
    \includegraphics[width=0.9\columnwidth]{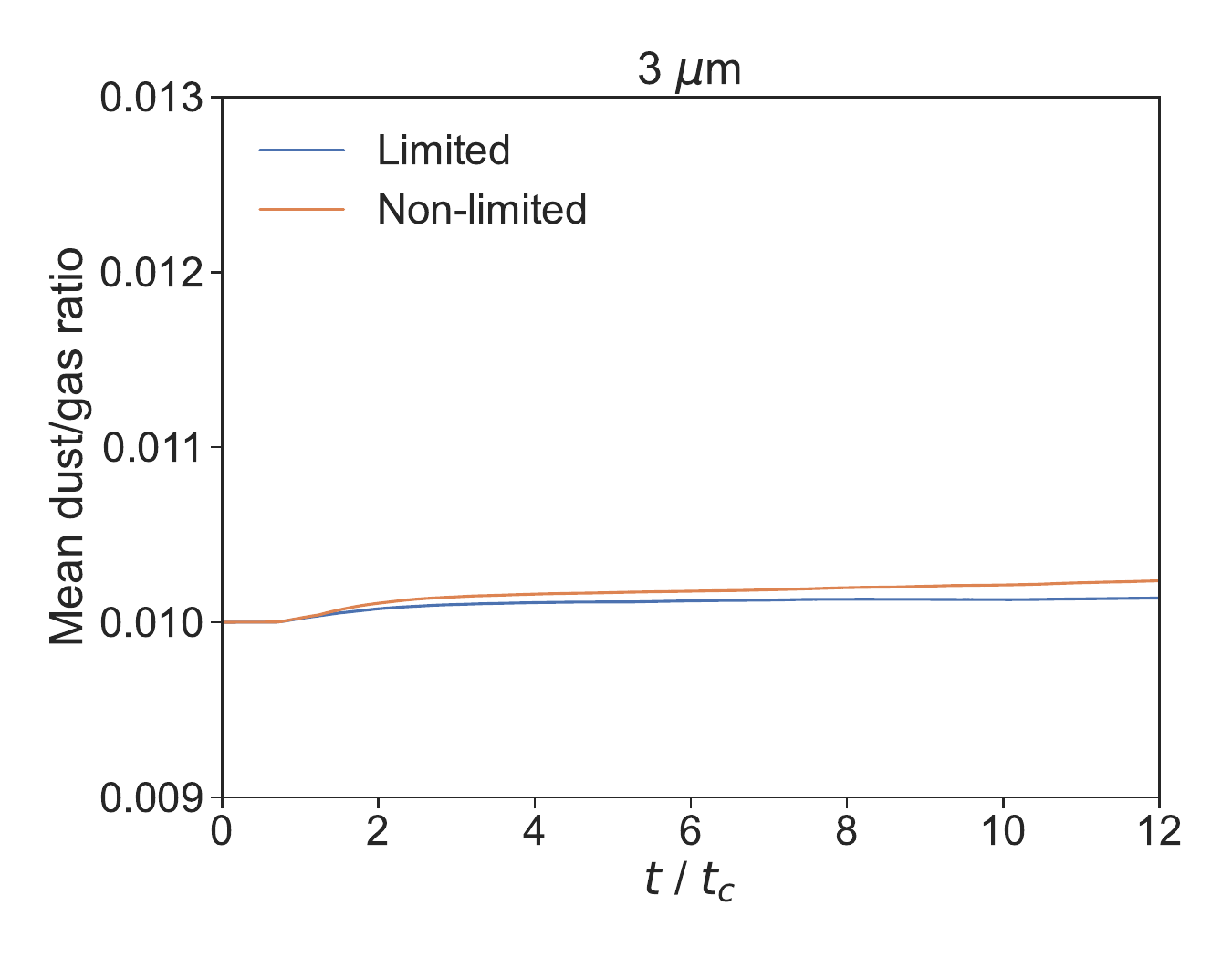} \\
    \includegraphics[width=0.9\columnwidth]{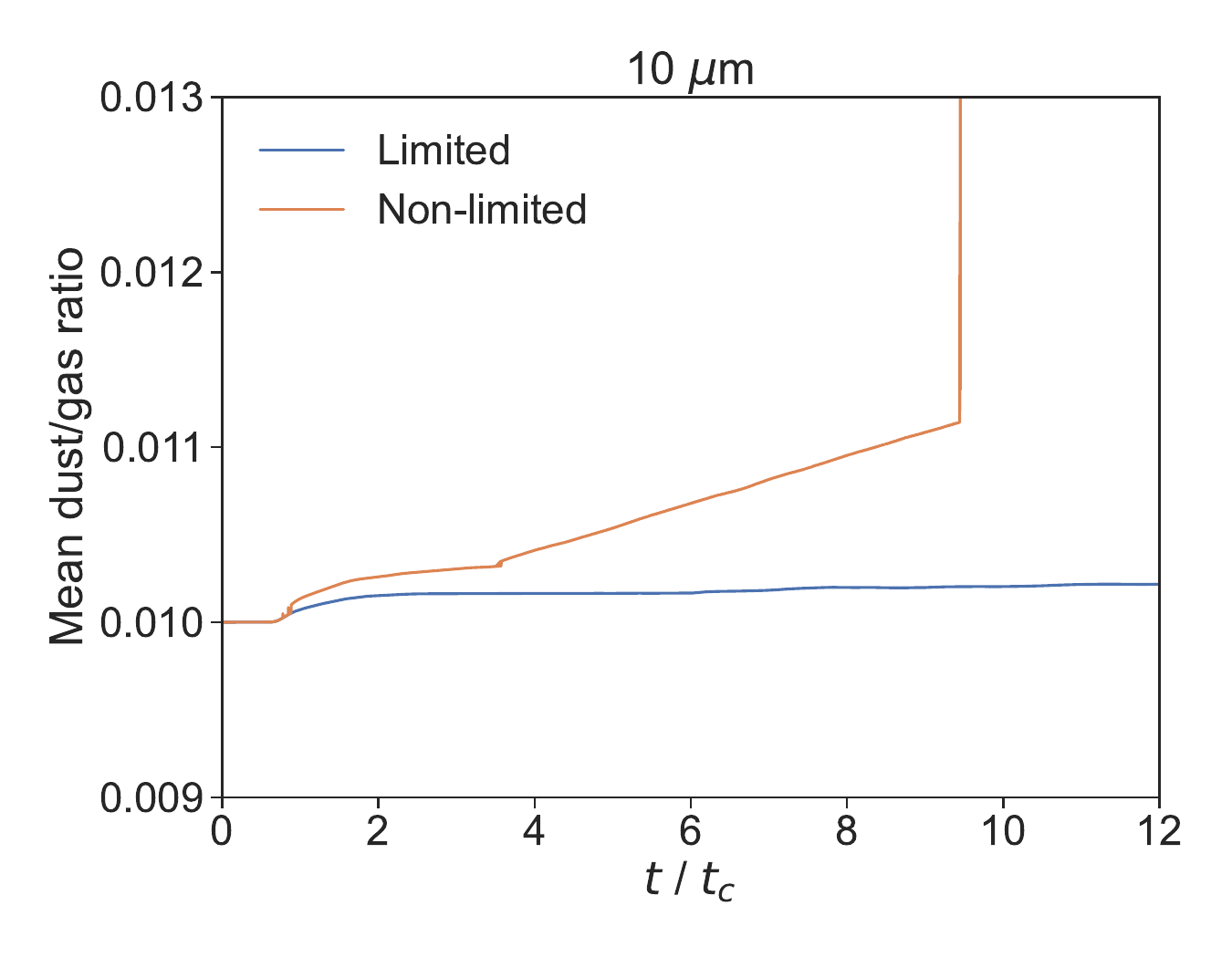} 
    \caption{Mean dust-to-gas ratio calculated on the particles in the one-fluid method with and without the stopping time limiter for 3 $\mu$m (top) and 10$\mu$m (bottom) dust grains in $256^3$ resolution.}
    \label{fig:mean_dtg_lnl}%
\end{figure}

Figure~(\ref{fig:mean_dtg_lnl}) shows the mean dust-to-gas ratio as calculated on the particles for results with and without the limiter. For 3 $\mu$m grains, there is a small, but noticeable increase in the mean dust-to-gas ratio from 1.014\% with the limiter to 1.024\% without at $t/t_{\rm c}=12$. However, there is a significantly greater difference between the limited and nonlimited results for 10 $\mu$m dust grains. With the limiter, the mean dust-to-gas ratio is 1.022\% at $t/t_{\rm c}=12$. Without the limiter, the mean dust-to-gas ratio steadily increases over time, reaching a mean dust-to-gas ratio of 1.14\% before experiencing numerical instability around $t/t_{\rm c} \approx 9.5$. This occurs due to the violation of dust mass conservation, in that, dust mass increases over time due to numerical error.

That numerical instability occurs without the limiter is a consequence of the dynamical nature of our dusty turbulence calculations. Low-density particles, that is, particles with St$_{\rm num} > 1$, can experience spurious increase in dust mass. For \citet{ballabio-etal18}, these errors occur on the periphery of a protoplanetary disk, which has little dynamical impact on the interior of the disk. However, in our calculations, the errors on these low-density particles quickly gets recycled into high-density filaments because the turbulence is driven for multiple turbulent crossing times. These errors compound over time. 

Though the limiter avoids catastrophic numerical failure, it does introduce some numerical effect on our results. The stopping time limiter has the same practical effect as reducing the dust grain size. As an extreme example, a one-fluid simulation with 1 cm grains would be expected to give results equivalent to a simulation with 10 cm grains because the same limited stopping time would be applied to all particles in both cases. 

\begin{figure}
    \centering
    \includegraphics[width=0.9\columnwidth]{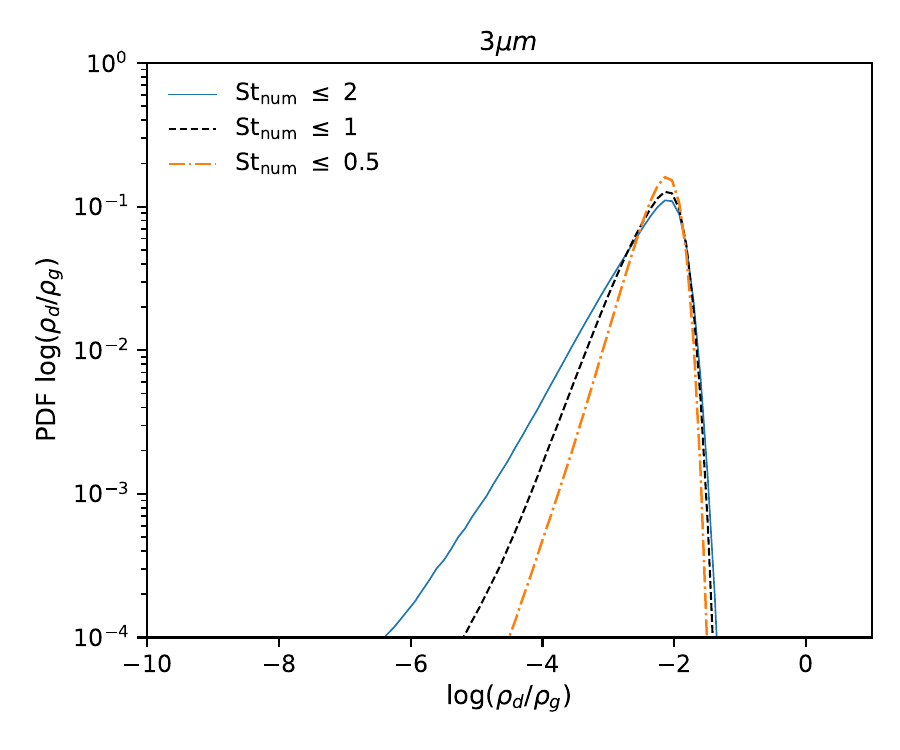} \\
    \includegraphics[width=0.9\columnwidth]{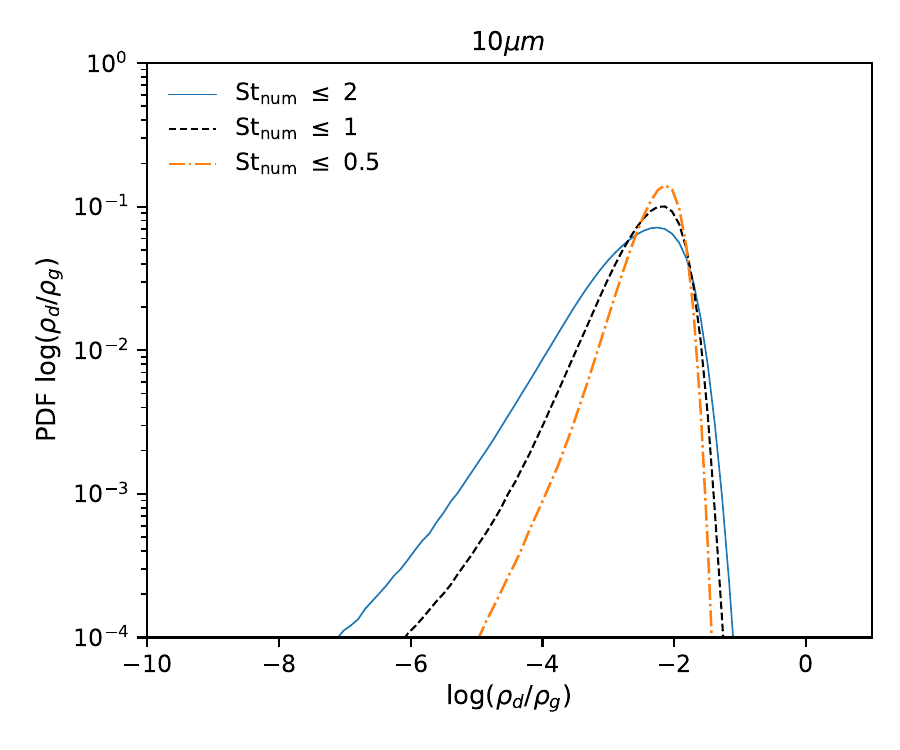} 
    \caption{Time-averaged volume-weighted PDFs of log$(\rho_{\rm d}/\rho_{\rm g})$ for 3 $\mu$m (top) and 10 $\mu$m (bottom) in $256^3$ resolution where the one-fluid stopping time limiter has been adjusted to restrict St$_{\rm num}$ $\leq$ 0.5 (orange) and $\leq$ 2 (blue) at $t/t_{\rm c}=5$. A stricter limiter further enhances the coupling of dust to gas, restricting the range of dust-to-gas ratios, while relaxing the limiter correspondingly broadens the distribution.}
    \label{fig:varying_limiter}
\end{figure}

The stopping time limiter restricts the numerical Stokes number to a maximum of one. We experimented with simulations where the limiter restricts the numerical Stokes number by a factor of 2 greater and smaller; that is, to St$_{\rm num} < 0.5$ and $< 2$, respectively. Figure~(\ref{fig:varying_limiter}) shows the PDFs of dust-to-gas ratios for 3 and 10 $\mu$m simulations with these limiter adjustments at $t/t_{\rm c}=5$. As expected, restricting the numerical Stokes number to a maximum of 0.5 tightens the width of the dust-to-gas ratio distributions, while relaxing the limiter to a maximum numerical Stokes number of two broadens the distributions. As a substantial fraction of particles are affected the limiter (see Figure~\ref{fig:stokes} and Section~\ref{sec:stokes_number}), this is a direct consequence of enforcing the dust to be more coupled to the gas in the former case and less coupled in the latter. Notably, relaxing the limiter still does not reproduce a dust-to-gas distribution matching the two-fluid results (Figure~\ref{fig:dtg_l2f}), but, as demonstrated, completely removing the limiter opens up numerical inaccuracies and instability for these grain sizes.

\section{Conclusions}
\label{sec:conclusion}

We performed SPH simulations of dusty, isothermal turbulence using 3 and 10 $\mu$m dust grains in initial conditions representative of cold, dense molecular clouds. Turbulent driving was applied to create and sustain supersonic $\mathcal{M}=10$ turbulence over 12 turbulent crossing times. 

Simulations were performed using the two- and one-fluid (with the diffusion approximation) methods. Both treat the dust as a continuous, pressureless fluid. For the two-fluid dust method, gas and dust are modeled using distinct sets of particles. These two types of particles are coupled together through a drag term. For the one-fluid method, a single species of particle is used that represents the mixture of gas and dust. Particles exchange dust between them, which changes the dust fraction of each particle. We use the terminal velocity approximation for the one-fluid method, in which case dust evolution acts like a diffusion process. The stopping time limiter of \citet{ballabio-etal18} was employed for numerical stability 

The goal was to compare the effect of dust solver on the response of the local dust-to-gas ratio due to turbulence. All calculations were performed using the Phantom SPH code. This ensured that all numerical details were consistent between calculations, aside from the choice of numerical dust solver. All calculations include the back-reaction of the dust on the gas.

Results were analyzed in the context of the numerical Stokes number, which is defined using the local dynamical time. This is the most relevant timescale under which to understand the numerical properties of the dust methods. The numerical Stokes numbers of the initial conditions for our calculations were St$_{\rm num} = 1.3$ and St$_{\rm num} = 4.3$ for 3 and 10 $\mu$m dust grains, respectively. This numerical Stokes number definition has a resolution dependence, such that it increases as the resolution of the calculation is increased. We specifically chose 3 and 10 $\mu$m dust grains so that the initial conditions were in the St$_{\rm num} \gtrsim 1$ regime in order to avoid excessively high resolution requirements for the two-fluid method \citep{lp12a}. Note, however, that the numerical Stokes number per particle ranges between 0.1 and 10 once turbulence is developed.

Our key findings are below.  

\begin{enumerate}

\item The PDFs of the gas density were lognormal, as is expected for isothermal, supersonic turbulence. No significant differences were found between the simulations with 3 and 10 $\mu$m dust grains, or between the one- and two-fluid methods. This additionally implies that the back-reaction of the dust on the gas had negligible impact on the gas density.

\item Dust density PDFs deviated from lognormal distributions. For both the two- and one-fluid methods, the PDFs peaked at values similar to what would be expected from the peak of the gas density PDF shifted by the initial dust-to-gas ratio. If the dust was tightly coupled to the gas, then lognormal distributions would be expected for the dust density PDFs matching the shape of gas density PDFs. Since they do not, this means that dust decouples from the gas in our calculations. The dust density PDFs for the two-fluid method is broader than for the one-fluid method, suggesting a greater degree of dust decoupling from the gas. The stopping time limiter in the one-fluid method is the primary reason for stronger coupling between gas and dust.

\item We found no evidence for orders of magnitude increase in the dust-to-gas ratio. PDFs of the dust-to-gas ratio peaked at approximately the mean dust-to-gas ratio of 1:100 for both grain sizes and both dust methods. However, the two-fluid dust-to-gas ratio PDFs were significantly broader than the one-fluid dust-to-gas ratio PDFs. The maximum dust-to-gas ratio for the one-fluid results was approximately a 3$\times$ and 5$\times$ increase over the initial ratio for 3 and 10 $\mu$m grains, respectively, whereas the two-fluid results reached maximum increased dust-to-gas ratios of approximately 30$\times$ and 39$\times$, respectively. This means the dust is more tightly coupled to the gas in the one-fluid results.

\item The one-fluid method stopping time limiter avoided spurious dust mass growth due to numerical inaccuracy (see Figure~\ref{fig:mean_dtg_lnl}). Calculations without the limiter exhibited a secondary peak in the dust-to-gas ratio PDFs at extremely low dust-to-gas ratios ($< 10^{-5}$). This low dust-to-gas ratio peak was in excess of the two-fluid PDFs. This behavior is even more pronounced for 10 $\mu$m dust grains compared to 3 $\mu$m dust grains, which is consequence of a larger fraction of particles exceeding St$_{\rm num} > 1$. This is in agreement with \citet{commercon-etal23}, who found that the terminal velocity approximation is well suited for $<$~10~$\mu$m dust grains inside dense filaments, but that it does not accurately capture dust dynamics in low-density regions. Though necessary to avoid numerical issues (or instability), the limiter introduces its own numerical artifacts manifesting as enhanced dust-gas coupling. 

\item There is reason to be cautious about the dust-to-gas ratio increases obtained with the two-fluid method. The right tail of the dust-to-gas ratio PDFs extended to higher values for the two-fluid method compared to the one-fluid method. However, this is the regime where the one-fluid method results are most accurate, as evidenced by the agreement of the right-hand tails of the PDF between the limited and nonlimited one-fluid results. This discrepancy raises concerns about potential spurious dust concentration arising from the two-fluid approach, especially for particles with St$_{\rm num} < 1$. Such spurious concentrations may occur due to nonadherence of the two-fluid resolution requirement, which, if not adequately met, has been show to lead to artificially high dust concentrations. These artifacts may result from insufficient resolution in capturing the coupled dynamics of dust and gas at smaller scales (inside dense filaments). Note that the turbulent driving force disperses filaments and artificial concentrations over a turbulent crossing time, preventing ongoing accumulation of dust over time. This aligns with the findings of \citet{commercon-etal23}, who showed that for tightly coupled dust grains, the apparent large dust-to-gas ratio variations are spurious, mostly because of artificial trapping in the high-density regions. 

\item The numerical Stokes number of a particle scales according to St$_{\rm num} \propto \rho^{-2/3}$ in SPH because the resolution is adaptive with respect to density. This means that changing the numerical resolution affects the numerical Stokes number, both in terms of increasing particle count as well as the adaptive particle resolution with respect to density. Note that even if the requirement around the numerical Stokes number for the one-fluid (St$_{\rm num} < 1$) and the two-fluid (St$_{\rm num} > 1$) methods is satisfied in the initial conditions, then it may not continued to be satisfied due to changing dynamical conditions. The numerical Stokes number increases inside dense filaments. The correct dust solver should be chosen with respect to the numerical Stokes number conditions encountered over the duration of the calculation.

\end{enumerate}

Our overall conclusion is that results obtained from simulations of dusty turbulence can be affected by the numerics of the dust solver. Both the two- and one-fluid with diffusion approximation can yield viable solutions, but the correct dust solver to use should be chosen based on the numerical Stokes number regime of the dust. If the simulation is fully in the St$_{\rm num} > 1$ domain, then it is expected that the two-fluid method should give accurate results. Conversely, the one-fluid method with diffusion approximation should achieve accurate results for a simulation fully in the St$_{\rm num} < 1$ domain. 

If the dust solver is used outside of the numerical Stokes regime that it is suited for, then numerical artifacts can appear as either (1) spurious dust concentrations (exaggerated dust-to-gas ratios) when using the two-fluid method, (2) enhanced coupling of dust and gas when using the one-fluid method with a stopping time limiter, or (3) artificial peaks at low dust-to-gas ratios and dust mass growth for the one-fluid method without limiter. 

It would be of interest to investigate the numerical properties of the full one-fluid formalism in the context of supersonic, dusty turbulence; that is, the one-fluid model without the terminal velocity approximation. This method should avoid the enhanced dust-gas coupling issue related to the stopping time limiter, permitting  accurate results for particles with St$_{\rm num} \gtrsim 1$ (i.e., low density). Being a one-fluid method, it should also not be susceptible to the spurious dust concentrations produced with the two-fluid method. The disadvantage would be increased computational cost to correctly resolve the dynamics in regions with St$_{\rm num} > 1$.

\begin{acknowledgments}
This project has been made possible in part by support provided by the Atlantic Canada Opportunities Agency (ACOA). We acknowledge the support of the Natural Sciences and Engineering Research Council of Canada (NSERC), grant number RGPIN/003786-2022. This research was enabled in part by support provided by ACENET (ace-net.ca) and the Digital Research Alliance of Canada (alliancecan.ca). Figures and analysis made use of Sarracen, a Python package for the analysis and visualization of SPH data \citep{sarracen}.
\end{acknowledgments}

\bibliographystyle{aasjournal}
\bibliography{bib}

%% This command is needed to show the entire author+affiliation list when
%% the collaboration and author truncation commands are used.  It has to
%% go at the end of the manuscript.
%\allauthors

%% Include this line if you are using the \added, \replaced, \deleted
%% commands to see a summary list of all changes at the end of the article.
%\listofchanges

\end{document}